\documentclass[article]{biometrika}

\usepackage{amsmath,amsfonts,mathtools,enumitem,stmaryrd,subfiles,gensymb,setspace,xcite}

\usepackage{times}
\usepackage{bm}
\usepackage{natbib}
\usepackage[plain,noend]{algorithm2e}

\makeatletter
\renewcommand{\algocf@captiontext}[2]{#1\algocf@typo. \AlCapFnt{}#2} 
\def\@algocf@capt@plain{top}
\renewcommand{\algocf@makecaption}[2]{%
  \addtolength{\hsize}{\algomargin}%
  \sbox\@tempboxa{\algocf@captiontext{#1}{#2}}%
  \ifdim\wd\@tempboxa >\hsize
    \hskip .5\algomargin%
    \parbox[t]{\hsize}{\algocf@captiontext{#1}{#2}}
  \else%
    \global\@minipagefalse%
    \hbox to\hsize{\box\@tempboxa}
  \fi%
  \addtolength{\hsize}{-\algomargin}%
}
\makeatother

\DeclareMathOperator{\Tr}{Tr}
\DeclareMathOperator{\E}{\mathbb{E}}
\DeclarePairedDelimiter\abs{\lvert}{\rvert}%
\DeclarePairedDelimiter\norm{\lVert}{\rVert}%
\DeclareMathOperator{\diag}{diag}

\usepackage{xr}
\makeatletter
\newcommand*{\addFileDependency}[1]{
  \typeout{(#1)}
  \@addtofilelist{#1}
  \IfFileExists{#1}{}{\typeout{No file #1.}}
}
\makeatother

\begin{document}

\jname{Submitted to Biometrika}
\jyear{2017}
\jvol{}
\jnum{}
\accessdate{}

\received{{\rm 2} January {\rm 2017}}
\revised{{\rm 1} April {\rm 2017}}


\markboth{Chris McKennan \and Dan Nicolae}{Biometrika style}

\title{Accounting for unobserved covariates with varying degrees of estimability in high dimensional biological data}

\author{Chris McKennan, Dan Nicolae}
\affil{Department of Statistics, University of Chicago, 5747 S. Ellis Avenue, Chicago, IL \email{cgm29@galton.uchicago.edu} \email{nicolae@galton.uchicago.edu}}

\maketitle

\begin{abstract}
An important phenomenon in high dimensional biological data is the presence of unobserved covariates that can have a significant impact on the measured response. When these factors are also correlated with the covariate(s) of interest (i.e. disease status), ignoring them can lead to increased type I error and spurious false discovery rate estimates. We show that depending on the strength of this correlation and the informativeness of the observed data for the latent factors, previously proposed estimators for the effect of the covariate of interest that attempt to account for unobserved covariates are asymptotically biased, which corroborates previous practitioners' observations that these estimators tend to produce inflated test statistics. We then provide an estimator that corrects the bias and prove it has the same asymptotic distribution as the ordinary least squares estimator when every covariate is observed. Lastly, we use previously published DNA methylation data to show our method can more accurately estimate the direct effect of asthma on methylation than previously published methods, which underestimate the correlation between asthma and latent cell type heterogeneity. Our re-analysis shows that the majority of the variability in methylation due to asthma in those data is actually mediated through cell composition.
\end{abstract}


\section{Introduction}
\label{section:introduction}
\indent There has been a rapid development of high throughput genetic and proteomic technologies to perform experiments to measure mRNA expression, protein expression and DNA methylation. However, analyzing these data has proven difficult because unmeasured factors that influence the observed data can have a detrimental impact on inference, especially when they are correlated with the variable of interest. For example, observed mRNA, proteomic and methylation data typically vary depending on reagent quality, laboratory temperature and the cellular composition of each sample \citep{Combat,LeekBatch,HousemanRef}, all of which are difficult or impossible to record. In this article, we show that, depending on how informative the data are for inferring the missing covariates, previous methods to correct for unobserved variables provide biased estimates for the effects of interest. We then provide an alternative method and prove one can do inference that is just as powerful as when the unobserved covariates are recorded, even when some of the unobserved covariates are difficult to estimate from the data.\par
\indent To develop some intuition for this problem, let $\bm{Y}_{p \times n}$ be the expression or methylation of $p$ units (i.e. genes, proteins or methylation sites) across $n$ samples. In a typical biological application, the goal is to estimate the effect of $d$ covariates of interest, whose observed values for each sample are given by the rows of $\bm{X}_{n \times d}$, on the expression or methylation at each of the $p$ units. In the presence of other unobserved variables $\bm{C}$ that may or may not influence $\bm{Y}$, a simple model would be
\begin{align}
\label{equation:Model0}
\bm{Y} &= \bm{B}\bm{X}^T + \bm{\Delta}\\
\bm{\Delta}_{p \times n} &= \bm{L}_{p \times K}\bm{C}_{n \times K}^T + \bm{E}_{p \times n}.
\end{align}
where $\bm{E}$ contains independent entries and identically distributed columns. When the effects due to $\bm{C}$ are non-zero, the naive ordinary least squares (OLS) estimator $\hat{\bm{B}}^{(\text{naive})} = \bm{Y}\bm{X}\left( \bm{X}^T\bm{X} \right)^{-1}$ is biased by $\bm{L}\left( \bm{\Omega}^{(OLS)} \right)^T$, where $\bm{\Omega}^{(OLS)} = \left( \bm{X}^T\bm{X} \right)^{-1} \bm{X}^T \bm{C}$ is the ordinary least squares coefficient estimate for the regression of $\bm{C}$ onto $\bm{X}$. The size of the bias is in part determined by the empirical effect of $\bm{X}$ on $\bm{C}$. For well designed experiments with large sample sizes, we would expect $\bm{\Omega}^{(OLS)}$ to be close to zero. However, when $p$ is large, the correlation between the $p$ rows of $\hat{\bm{B}}^{(\text{naive})}$ induced by the unobserved covariates tends to obfuscate inference, even for large sample sizes \citep{lfdr,CorrZScores}. There are other cases where $\bm{\Omega}^{(OLS)}$ will not be close to zero no matter how large the sample size is. For example, if $\bm{X}$ were a measurement of environment or disease status and $\bm{Y}$ were DNA methylation, then unmeasured cellular heterogeneity may vary with $\bm{X}$, which would subsequently alter measured methylation \citep{Michelle,Cell}.\par 
\indent There have been a number of methods proposed to solve this problem \citep{SVA,RUV,Houseman,LEAPP,BiometrikaConfounding,Fan1,CATE}.  \citet{SVA} try to identify units where the effect due to $\bm{X}$ is 0 and do factor analysis on only those factors to estimate $\bm{C}$. The method proposed in \citet{RUV} is very similar to that of \citet{SVA}, except they assume the practitioner has prior knowledge of a subset of the $p$ units whose response does not depend on $\bm{X}$. While this performs well when such a subset is known, it is rare for practitioners to have such strong prior information. In \citet{Houseman,LEAPP}, the authors use factor analysis to estimate $\bm{L}$ and use the estimate to remove the bias in the naive ordinary least squares estimate for $\bm{B}$. While the authors of these two articles show their methods perform well on selected data sets, they do not provide sufficient theory to justify inference using their estimators. Lastly, \citet{BiometrikaConfounding} provided conditions for which the estimators of individual rows of $\bm{B}$ are consistent, but did not provide any theory necessary to perform inference.\par 
\indent Recently, \citet{Fan1,CATE} proposed methods that estimated first $\bm{L}$ using factor analysis on the residuals $\bm{Y} - \hat{\bm{B}}^{(\text{naive})} \bm{X}^T$, estimated $\bm{\Omega}^{(OLS)}$ by regressing $\hat{\bm{B}}^{(\text{naive})}$ onto $\hat{\bm{L}}$ and then estimated $\bm{B}$ by removing the estimated bias $\hat{\bm{L}} \hat{\bm{\Omega}}^{(OLS)T}$ from $\hat{\bm{B}}^{(\text{naive})}$. \citet{Fan1} proved that when $\bm{C}$ was independent of $\bm{X}$, their estimate for the false discovery rate was asymptotically correct and \citet{CATE} proved their estimates for a single row of $\bm{B}$ (i.e. the effects for a single unit) had the same asymptotic distribution as when $\bm{C}$ was known. However, it has been shown that these methods tend to inflate and bias test statistics in practice \citep{BeyondCATE}. One source of this discrepancy between theory and practice in both articles is the critical assumption that all $K$ of the eigenvalues of $p^{-1} P_{\bm{X}}^{\perp}\bm{C} \bm{L}^T \bm{L}\bm{C}^T P_{\bm{X}}^{\perp}$ are on the order of the number of samples, $n$, where $P_{\bm{X}}^{\perp}$ is the orthogonal projection matrix for the orthogonal complement of $\bm{X}$. That is, they assumed the unobserved variable's effects were easily estimated from the data. However, this is rarely the case in real data applications, especially in methylation data when unmeasured cellular heterogeneity is correlated with the covariate of interest \citep{Cell}. The purpose of this article is therefore to fill this gap in the literature by studying this problem when the data may or may not be informative for the unobserved covariates.\par 
\indent The remainder of the paper is organized as follows: we first introduce the model for the data in Section \ref{section:methods} and describe our estimation procedure and the conditions each step must satisfy so we can perform accurate inference. We then make our first contribution in Section \ref{subsection:AsymptoticBias}, where we prove that if the data are not informative for the unobserved covariates (i.e. the eigenvalues of $p^{-1} P_{\bm{X}}^{\perp}\bm{C} \bm{L}^T \bm{L}\bm{C}^T P_{\bm{X}}^{\perp}$ fall below a certain threshold), previously proposed estimates for $\bm{B}$ are asymptotically biased. We make our second and most important contribution in Section \ref{CorrectingBias}, where we provide a bias-corrected estimator for the effect of $\bm{X}$ on each unit's expression or methylation. We then prove its asymptotic distribution is the same as the ordinary least squares estimator when $\bm{C}$ is observed, regardless of how informative the data are for the unmeasured covariates. Lastly, we use simulated and recently published DNA methylation data to show our method can better account for latent covariates than the leading competitors, which can greatly alter the biological interpretation of the data. The proofs of all propositions, lemmas, theorems and corollaries are given in the Supplement.

\section{Models, motivation and intuition}
\label{section:methods}
\subsection{A model for the data}
\label{subsection:Datamodel}
We assume the data $\bm{y}_i \in \mathbb{R}^p$, $i = 1, 2, \ldots, n$, are independent and we define the data matrix $\bm{Y} = \begin{bmatrix}
\bm{y}_1 & \cdots & \bm{y}_n
\end{bmatrix} \in \mathbb{R}^{p \times n}$. For example, if $\bm{Y}$ were DNA methylation data, $\left\lbrace \bm{y}_i \right\rbrace_{i=1}^n$ is the measured DNA methylation across $p$ cytosines for samples $i = 1, 2, \ldots, n$. For any matrix $\bm{G} \in \mathbb{R}^{n \times m}$, we define $P_{\bm{G}}$ and $P_{\bm{G}}^{\perp}$ to be the orthogonal projection matrices that project vectors in $\mathbb{R}^n$ onto the image of $\bm{G}$ and the orthogonal complement of $\bm{G}$, respectively. Let $\bm{X} \in \mathbb{R}^{n \times d}$ be the covariate(s) of interest and $\bm{B} = \begin{bmatrix}
\bm{\beta}_1 & \cdots & \bm{\beta}_p
\end{bmatrix}^T \in \mathbb{R}_{p \times d}$ their corresponding effects across all $p$ variables. We also define an additional covariate matrix $\bar{\bm{C}} \in \mathbb{R}^{n \times K}$ and $\bar{\bm{L}} = \begin{bmatrix}
\bar{\bm{\ell}}_1 & \cdots & \bar{\bm{\ell}}_p
\end{bmatrix}^T \in \mathbb{R}^{p \times K}$ their corresponding effects. We will assume that $\bar{\bm{C}}$ is unobserved but $K$ is known. Of course, $K$ is rarely known in true data applications. While we acknowledge that estimating $K$ is a non-trivial problem, there is a large body of work devoted to estimating it \citep{SVA,RUV,EigenDiff,bcv}. We discuss how different values of $K$ affect our downstream estimates in Sections \ref{section:DataAnalysis} and \ref{Discussion}. The full model for the data is then taken to be
\begin{align}
\label{equation:Y.model}
\bm{Y}_{p \times n} &= \bm{B}_{p \times d}\bm{X}_{n \times d}^T + \bar{\bm{L}}_{p \times K}\bar{\bm{C}}_{n \times K}^T + \bm{E}_{p \times n} \quad \text{where $\bm{E}_{p \times n} \sim MN_{p \times n}\left( \bm{0}, \bm{\Sigma}_{p \times p}, I_n \right)$}\\
\label{equation:Y.rho}
\rho &= \frac{1}{p}\Tr\left( \bm{\Sigma} \right).
\end{align}
We then make the following technical assumptions about $\bm{X}$, $\bar{\bm{C}}$ and $\bm{E}$:
\begin{assumption}
\label{assumption:OLS}
\begin{enumerate}[label=(\alph*)]
\item $\bm{X}$ is a non-random, full rank matrix with $\lim_{n \to \infty} \frac{1}{n} \bm{X}^T \bm{X} \to \bm{\Sigma}_X \succ \bm{0}$.
\item $\bar{\bm{C}} = \bm{X}\bar{\bm{\Omega}} + \bar{\bm{\Xi}}$ where 
\begin{align*}
\frac{1}{n}\bar{\bm{\Xi}}^T \bar{\bm{\Xi}} \stackrel{P}{\to} \bar{\bm{\Psi}} \succ \bm{0}.
\end{align*}
\item $\bar{\bm{C}}$ is independent of $\bm{E}$.
\item $\bm{\Sigma} = \text{diag}\left( \sigma_1^2, \ldots, \sigma_p^2 \right)$ and $\sigma_g^2 \in \left[ c_1^{-1}, c_1 \right]$ $\forall g = 1, \ldots, p$ and some constant $c_1 > 0$ that does not depend on $n$ or $p$.
\end{enumerate}
\end{assumption}
Items (a), (b) and (c) are standard linear modeling assumptions and (d) simply bounds the residual variances. Lastly, we define the matrix $\bm{A} \in \mathbb{R}^{n \times (n-d)}$ whose columns form an orthonormal basis for $\text{ker}\left( \bm{X}^T \right)$ and
\begin{align}
\label{equation:DefineOmegaOLS}
\bar{\bm{\Omega}}^{(OLS)} = \bar{\bm{\Omega}} + \left( \bm{X}^T \bm{X} \right)^{-1}\bm{X}^T\bar{\bm{\Xi}}
\end{align}
to be the coefficients from the regression $\bar{\bm{C}}$ onto $\bm{X}$. Note that $\norm{\bar{\bm{\Omega}}^{(OLS)} - \bar{\bm{\Omega}}}_2 = \mathcal{O}_P(1)$ as $n \to \infty$ by items (a) and (b).\\
\indent A more general model for $\bm{Y}$ would be
\begin{align}
\label{equation:NuisanceCov}
\bm{Y}_{p \times n} = \bm{B}_{p \times d}\bm{X}_{n \times d}^T + \bm{M}_{p \times r}\bm{Z}_{n \times r}^T + \bar{\bm{L}}_{p \times K}\bar{\bm{C}}_{n \times K}^T + \bm{E}_{p \times n}
\end{align}
where $\bm{Z}$ are observed nuisance factors whose effects we are not interested in (e.g. the intercept and other biological and/or technical covariates). We can get back to model \eqref{equation:Y.model} by simply multiplying $\bm{Y}$ on the right by a matrix whose columns form an orthonormal basis for $\text{ker}\left( \bm{Z}^T \right)$. Therefore, we work exclusively with \eqref{equation:Y.model} and assume any nuisance factors have already been rotated out.\\
\indent Using a technique developed in \cite{LEAPP}, we break $\bm{Y}$ into two independent pieces:
\begin{align}
\label{equation:Y1.model}
\bm{Y}_1 &= \bm{Y}\bm{X}\left( \bm{X}^T \bm{X} \right)^{-1} = \bm{B} + \bar{\bm{L}}\bar{\bm{\Omega}}^{(OLS)T} + \bm{E}_1\\
\label{equation:Y2.model}
\bm{Y}_2 &= \bm{Y}\bm{A} = \bar{\bm{L}}\bar{\bm{C}}_2^T + \bm{E}_2, \quad \bar{\bm{C}}_2 = \bm{A}^T \bar{\bm{C}}
\end{align}
where $\bm{E}_1 \sim MN_{p \times d}\left( \bm{0}, \bm{\Sigma}, \left(\bm{X}^T \bm{X}\right)^{-1} \right)$ and $\bm{E}_2 \sim MN_{p \times (n-d)}\left( \bm{0}, \bm{\Sigma}, I_{(n-d)} \right)$ are independent because $\bm{A}^T \bm{X} = \bm{0}_{(n-d) \times d}$. Note that $\bm{Y} = \bm{Y}_1 \bm{X}^T + \bm{Y}_2\bm{A}^T = \bm{Y}P_{\bm{X}} + \bm{Y}P_{\bm{X}}^{\perp}$ is a partition of $\bm{Y}$ into the variability due to $\bm{X}$ and the corresponding residuals. In what follows, we will use $\bm{Y}_2$ to estimate $\bm{L}$ and $\bm{\Sigma}$ and then plug these estimates into the mean and variance of $\bm{Y}_1$ to estimate $\bm{B}$, just as one would do in ordinary least squares.\par
\indent As will become apparent in Section \ref{Properties}, an important feature of equations \eqref{equation:Y2.model} is the magnitude of $\bar{\bm{L}}\bar{\bm{C}}_2^T$ determines how difficult it is to separate the variability in $\bm{Y}$ due to $\bm{X}$ from the variability due to $\bm{C}$. We say the data are informative for the confounders if the effect $\bar{\bm{L}}\bar{\bm{C}}_2^T$ is strong, and not informative if it is weak. This is very closely related to the size of the effect $\bar{\bm{L}}\bar{\bm{\Omega}}^{(OLS)T}$ from equation \eqref{equation:Y1.model}, since if this is large, it generally means $\bar{\bm{L}}\bar{\bm{C}}_2^T$ is weak. We define the informativeness precisely in section \ref{Properties}.\\
\indent Since we are only interested in estimating $\bm{B}$ and not the true value of $\bar{\bm{L}}$, we may modify $\bar{\bm{L}}$ and $\bar{\bm{C}}$ in any way we please, with the restriction that the product $\bar{\bm{L}}\bar{\bm{C}}^T$ remain the same. Therefore,
\begin{equation*}
\bar{\bm{L}}\bar{\bm{C}}^T = \left( \bar{\bm{L}}\hat{\bar{\bm{\Psi}}}^{1/2} \right) \left( \hat{\bar{\bm{\Psi}}}^{-1/2}\bar{\bm{C}}^T \right) = \bm{L}\bm{C}^T \text{ for $\hat{\bar{\bm{\Psi}}} = \frac{1}{n-d}\bar{\bm{C}}^T P_{X}^{\perp}\bar{\bm{C}} = \frac{1}{n-d}\bar{\bm{C}}_2^T \bar{\bm{C}}_2$}
\end{equation*}
and
\begin{equation*}
\frac{1}{n-d} \bm{C}_2^T \bm{C}_2 = \frac{1}{n-d}\hat{\bar{\bm{\Psi}}}^{-1/2} \bar{\bm{C}}_2^T \bar{\bm{C}}_2 \hat{\bar{\bm{\Psi}}}^{-1/2} = I_{K}.
\end{equation*}
We then replace $\bar{\bm{C}}$, $\bar{\bm{\Omega}}$ and $\bar{\bm{L}}$ with their standardized equivalents:
\begin{subequations}
\label{equations:Standardize}
\begin{align}
\bm{C} &= \bar{\bm{C}} \hat{\bar{\bm{\Psi}}}^{-1/2}\\
\bm{\Omega} &= \bar{\bm{\Omega}} \hat{\bar{\bm{\Psi}}}^{-1/2}\\
\bm{\Omega}^{(OLS)} &= \bar{\bm{\Omega}}^{(OLS)} \hat{\bar{\bm{\Psi}}}^{-1/2}\\
\bm{L} &= \bar{\bm{L}}\hat{\bar{\bm{\Psi}}}^{1/2}
\end{align}
\end{subequations}
where now
\begin{align*}
\frac{1}{n-d} \bm{C}_2^T \bm{C}_2 = I_K.
\end{align*}
Under this restriction, $\bm{L}$ and $\bm{C}$ are determined up to a rotation matrix. We may therefore assume that $\bm{L}^T \bm{L}$ is diagonal with decreasing elements. We will refer to this properly scaled and rotated $\bm{L}$ as the standardized confounding effects. We now present an additional set of assumptions that will be important for the remainder of the paper.
\begin{assumption}
\label{assumption:Set1}
\begin{enumerate}[label=(\alph*)]
\item $\frac{n-d}{p}\bm{L}^T \bm{L} = \text{diag}\left( \lambda_1, \ldots, \lambda_K \right)$ where $c_2^{-1} \leq \lambda_k \leq c_2 n$ and $\frac{\lambda_k - \lambda_{k+1}}{\lambda_k} \geq c_2^{-1}$ for all $k = 1, \ldots, K$ ($\lambda_{K+1} := 0$) and some constant $c_2 > 0$. Note that $\lambda_1, \ldots, \lambda_K$ are functions of $n$ and $p$.\label{item:Lambda}
\item The magnitude of the entries of $\bm{L}$ are uniformly bounded by some constant $c_3 > 0$.\label{item:BoundL}
\item $\frac{1}{n-d}\bm{C}_2^T \bm{C}_2 = I_K$.\label{item:Psi}
\item $\frac{n}{p} < 1$ for all $n, p$ and $\frac{n^{3/2}}{p \lambda_K} \to 0$ as $n,p \to \infty$.\label{item:SampleSize}
\end{enumerate}
\end{assumption}
Items \ref{item:Lambda} and \ref{item:Psi} are without loss of generality using arguments presented above. Item \ref{item:Lambda} also gives a proper definition of the informativeness of each confounding component: the larger $\lambda_k$, the more informative the data are for the $k^{\text{th}}$ confounding component. Previous work has only considered the case when that data are as informative as possible, i.e. $\lambda_k \asymp n$ for all $k = 1, \ldots, K$ \citep{Bai,CATE,Fan1}. Further, item \ref{item:SampleSize} is the sufficient condition given in \citet{CATE} to perform accurate inference on their estimate for $\bm{\beta}_g$ when $\lambda_k \asymp n$. However, there has been little work done when $\lambda_k = o(n)$ for some or all of the $K$ latent factors. 

\subsection{Intuition and overview of estimation}
\label{OLSReview}
Here we provide a brief overview of how we estimate and do inference on $\bm{\beta}_g$ when $\bm{C}$ is unobserved using intuition from ordinary least squares. When both $\bm{X}$ and $\bm{C}$ are observed, there is a natural way to use $\bm{Y}_1$ and $\bm{Y}_2$ from \eqref{equation:Y1.model} and \eqref{equation:Y2.model} to obtain $\hat{\bm{B}}^{OLS}$, the ordinary least squares estimate for $\bm{B}$. This procedure will give us insight into how we should tackle to problem when $\bm{C}$ is unobserved. The first step is to estimate $\bm{L}$ using $\bm{Y}_2$:
\begin{equation*}
\hat{\bm{L}}^{OLS} = \bm{Y}_2 \bm{C}_2\left( \bm{C}_2^T \bm{C}_2 \right)^{-1} \sim \bm{L} + \bm{Z}_2, \quad \bm{Z}_2 \sim MN_{p \times K}\left( \bm{0}, \bm{\Sigma}, \frac{1}{n-d} I_K \right)
\end{equation*}
where $\bm{Z}_2$ is independent of $\bm{E}_1$. We then compute $\bm{\Omega}^{(OLS)}=\left( \bm{X}^T\bm{X} \right)^{-1} \bm{X}^T \bm{C}$ and $\hat{\sigma}_{g,OLS}^2$, the unbiased ordinary least squares estimate for $\sigma_g^2$. The estimate for $\bm{\beta}_g$ is then
\begin{equation*}
\hat{\bm{\beta}}_g^{OLS} = \bm{y}_{1_g} - \bm{\Omega}^{(OLS)}\hat{\bm{\ell}}_g^{OLS}  = \bm{\beta}_g + \bm{\Omega}^{(OLS)}\left( \bm{\ell}_g - \hat{\bm{\ell}}_g^{OLS} \right) + \bm{e}_{1_g}
\end{equation*}
where $\bm{y}_{1_g}$ and $\bm{e}_{1_g}$ are the $g^{\text{th}}$ rows of $\bm{Y}_1$ and $\bm{E}_1$. Since $\bm{e}_{1_g}$ and $\hat{\bm{\ell}}_g^{OLS}$ are independent and $\hat{\sigma}_{g,OLS}^2 \stackrel{P}{\to} \sigma_g^2$, 
\begin{align*}
\frac{n^{1/2}}{\hat{\sigma}_g}\left( \hat{\bm{\beta}}_g - \bm{\beta}_g \right) \sim N_d\left( 0, \left( n^{-1}\bm{X}^T \bm{X} \right)^{-1} + \bm{\Omega}^{(OLS)} \bm{\Omega}^{(OLS)T} \right) + o_P(1)
\end{align*}
\par
\indent Even though this procedure is straightforward, it gives us insight into how we should proceed in estimating $\bm{B}$ when we do not observe the confounders $\bm{C}$.
\begin{algo}
\label{procedure:Est}
Suppose we observed $\bm{X}$ but not $\bm{C}$. The following is a general algorithm to estimate and do inference on $\bm{B}$ that mimics the OLS procedure above:
\begin{enumerate}
\item Use $\bm{Y}_2$ to obtain $\hat{\bm{L}}$ and $\hat{\bm{\Sigma}}$ in such a way that $\hat{\bm{\ell}}_g$ has the same asymptotic distribution as if $\bm{C}$ were observed and $\hat{\sigma}_g^2 \stackrel{P}{\to} \sigma_g^2$. We will show that we can do this with principal components analysis.
\item Use $\hat{\bm{L}}$ and $\bm{Y}_1$ to estimate $\bm{\Omega}^{(OLS)}$ in such a way that $n^{1/2}\left( \hat{\bm{\Omega}}^{(OLS)} - \bm{\Omega}^{(OLS)} \right) =o_P(1)$. This can be done by regressing $\bm{Y}_1$ onto $\hat{\bm{L}}$, under proper assumptions.
\item Set $\hat{\bm{\beta}}_g = \bm{y}_{1_g} - \hat{\bm{\Omega}}^{(OLS)} \hat{\bm{\ell}}_g$. The asymptotic distribution for $\hat{\sigma}_g^{-1}\hat{\bm{\beta}}_g$ is then
\begin{align*}
\frac{n^{1/2}}{\hat{\sigma}_g}\left( \hat{\bm{\beta}}_g - \bm{\beta}_g \right) &=  \bm{\Omega}^{(OLS)} \frac{n^{1/2}}{\hat{\sigma}_g} \left( \bm{\ell}_g - \hat{\bm{\ell}}_g \right) + \frac{n^{1/2}}{\hat{\sigma}_g}\bm{e}_{1_g} + \frac{n^{1/2}}{\hat{\sigma}_g}\left( \bm{\Omega}^{(OLS)} - \hat{\bm{\Omega}}^{(OLS)} \right)\hat{\bm{\ell}}_g\\ 
&\sim N_d\left( 0, \left( n^{-1}\bm{X}^T \bm{X} \right)^{-1} + \bm{\Omega}^{(OLS)} \bm{\Omega}^{(OLS)T} \right) + o_P(1)
\end{align*}
which is the exact asymptotic distribution as the ordinary least sqaures estimator $\hat{\bm{\beta}}_g^{OLS}$ when $\bm{C}$ is observed. 
\end{enumerate}
\end{algo}
The first goal of the paper is to show that under certain conditions, previously proposed estimators for $\bm{L}$ and $\bm{\Omega}^{(OLS)}$ when $\bm{C}$ is unobserved do not satisfy the condition in 2 of Algorithm \ref{procedure:Est}, meaning the asymptotic distribution of $\hat{\bm{\beta}}_g$ is not as given above. In fact, when the data are not informative for the confounding, previously proposed estimators for $\bm{\Omega}^{(OLS)}$ are asymptotically biased, with the bias getting more severe as the signal strength $\bm{L}\bm{C}_2^T$ gets weaker. We will then provide estimators for $\bm{L}$, $\bm{\Sigma}$ and $\bm{\Omega}^{(OLS)}$ and sufficient assumptions for them to satisfy conditions 1 and 2 above, even when the data are not informative for $\bm{C}$.

\section{Estimation with unobserved covariates}
\label{Properties}
\subsection{There is an asymptotic bias in standard estimation procedures}
\label{subsection:AsymptoticBias}
\indent Recall from algorithm \ref{procedure:Est} that we first need an to estimate $\bm{L}$ and $\bm{\Sigma}$ using $\bm{Y}_2$. Let
\begin{align*}
\frac{1}{n-d} \bm{Y}_2 \bm{Y}_2^T = \bm{F}_{p \times (n-d)}\text{diag}\left( \gamma_1^2, \ldots, \gamma_{n-d}^2 \right) \bm{F}_{p \times (n-d)}^T
\end{align*}
be the eigen-decomposition of the empirical covariance matrix. We then define the estimates
\begin{align}
\label{equation:Lhat}
\hat{\bm{L}} &= \bm{F}[,1:K] \text{diag}\left( \gamma_1, \ldots, \gamma_{K} \right) \text{ and $\hat{\lambda}_k = \frac{n-d}{p} \gamma_k^2$}\\
\label{equation:C2hat}
\hat{\bm{C}}_2 &= \bm{Y}_2^T\hat{\bm{L}} \left( \hat{\bm{L}}^T \hat{\bm{L}} \right)^{-1} \text{ and $\hat{\sigma}_g^2 = \frac{1}{n-d-K} \bm{y}_{2_g}^T P_{\hat{\bm{C}}_2}^{\perp}\bm{y}_{2_g}$}.
\end{align}
We then regress $\bm{Y}_1$ onto the noisy design matrix $\hat{\bm{L}}$ to get $\bm{\Omega}^{(OLS)}$, as is done in previously proposed procedures \citep{LEAPP,CATE,Fan1}
\begin{align}
\label{equation:Omega.biased}
\hat{\bm{\Omega}}^{(OLS)} = \bm{Y}_1^T \hat{\bm{L}} \left( \hat{\bm{L}}^T \hat{\bm{L}} \right)^{-1}.
\end{align}
The limitation of this procedure is when the standardized confounding effects are small, the residual $\hat{\bm{R}} = \hat{\bm{L}} - \bm{L}$ is relatively large in comparison to $\bm{L}$. If we consider the extreme when $\bm{L} = \bm{0}$, then the regression coefficients from the regression $\bm{Y}_1 \sim \hat{\bm{L}}$ should be very close to 0, since $\hat{\bm{L}} = \hat{\bm{R}}$ is independent of $\bm{Y}_1$. Therefore, the smaller the standardized confounding effects are, the more we would expect our naive estimate $\hat{\bm{\Omega}}^{(OLS)}$ to shrink closer to 0. We can formalize this discussion with the following proposition
\begin{proposition}
\label{proposition:BiasedOmega}
Suppose assumptions \ref{assumption:OLS} and \ref{assumption:Set1} hold.  Assume the eigenvalues $\lambda_k$ are of the same order, i.e. $\frac{\lambda_1}{\lambda_K} \leq c_4$ for some $c_4 > 0$, and the primary effect $\bm{B} = \bm{0}$. If we estimate $\bm{L}$ using \eqref{equation:Lhat} and $\bm{\Omega}^{(OLS)}$ using \eqref{equation:Omega.biased}, then
\begin{align}
\label{equation:ell.BiasedOmega}
\norm{\hat{\bm{\ell}}_g - \bm{\ell}_g}_2  =  \mathcal{O}_P\left( n^{-1/2} \right)
\end{align}
and
\begin{align}
\label{equation:Omega.BiasedOmega}
\norm{\hat{\bm{\Omega}}^{(OLS)} - \bm{\Omega}^{(OLS)}\text{diag}\left( \frac{\lambda_1}{\lambda_1 + \rho}, \ldots, \frac{\lambda_K}{\lambda_K  + \rho} \right)}_2 = o_P\left( n^{-1/2} \right)
\end{align}
where $\rho$ is defined in \eqref{equation:Y.rho}.
\end{proposition}
The consequence of this result is the naive estimator for $\bm{\Omega}^{(OLS)}$ given by \eqref{equation:Omega.biased} is asymptotically biased, with the magnitude of the bias becoming more significant as the signal strength of the standardized confounding effects decreases. Specifically, the asymptotic distribution for $n^{1/2}\left( \hat{\bm{\beta}}_g - \bm{\beta}_g \right)$ does not have mean 0, and instead is centered around something of magnitude $n^{1/2}/ \lambda_K$, which can be large depending on how informative the data are for the confounding. This means that if we assume
\begin{align*}
\frac{n^{1/2}}{\hat{\sigma}_g}\left( \hat{\bm{\beta}}_g - \bm{\beta}_g \right) \approx N_d\left( 0, \left( n^{-1}\bm{X}^T \bm{X} \right)^{-1} + \hat{\bm{\Omega}}^{(OLS)} \left( \hat{\bm{\Omega}}^{(OLS)} \right)^T \right),
\end{align*}
and $\lambda_K = O\left( n^{1/2} \right)$, we tend to introduce type I errors.\par
\indent Equation \eqref{equation:Omega.BiasedOmega} also implies $p/n$ is the lower limit of confounding detection. If the standardized confounding effect signal falls below $p/n$, \eqref{equation:Omega.BiasedOmega} says that $\hat{\bm{\Omega}}^{(OLS)} = \bm{0}$ and we have no hope of correcting for confounding. However, it would be a mistake to think that increasing the sample size while keeping the number of sites $p$ constant should make confounder correction more difficult. In fact, it is the opposite. In most data, $p$ and $\bm{L}$ are fixed, meaning the eigenvalues $\lambda_k$ grow linearly with the sample size $n$. Therefore, the bias actually decays as the sample size increases (assuming the number of confounding variables $K$ remains fixed), which is exactly what one would expect.\par 
\indent The estimators described in \cite{LEAPP,CATE,Fan1} do not estimate $\bm{\Omega}^{(OLS)}$ using equation \eqref{equation:Omega.biased}. If $\bm{B} = \bm{0}$, then the ordinary least squares estimator in \eqref{equation:Omega.biased} is a reasonable choice. However, one might expect that when $\bm{B} \neq \bm{0}$, the estimator in \eqref{equation:Omega.biased} may be ``contaminated" by the non-zero $\bm{B}$. Therefore, the authors of \cite{LEAPP,CATE,Fan1} use more robust estimators to alleviate contamination by a sparse, non-zero $\bm{B}$. While their estimators for $\bm{\Omega}^{(OLS)}$ are not shrunk in the exact way \eqref{equation:Omega.BiasedOmega} predicts, simulations in section \ref{subsection:Simulations} show the shrinkage is just as substantial.

\subsection{Correcting the asymptotic bias}
\label{CorrectingBias}
\indent Now that we have shown how the bias can compromise inference, we provide bias-corrected estimators $\hat{\bm{\beta}}_g^{bc}$ and $\hat{\sigma}_g^2$ for the main effect $\bm{\beta}_g$ and variance $\sigma_g^2$ and prove that the asymptotic distribution of $\hat{\sigma}_g^{-1}\hat{\bm{\beta}}_g^{bc}$ is the same as if we had observed the latent covariates $\bm{C}$, even when the data are not informative for $\bm{C}$. \par 
\indent For the rest of this section, we will assume the data $\bm{Y}$ have been generated according to \eqref{equation:Y.model} and we estimate $\bm{L}$ and $\lambda_k$ according to \eqref{equation:Lhat} and $\bm{\Sigma}$ according to \eqref{equation:C2hat}. The following two lemmas will be important in deriving the asymptotic distribution of $\hat{\sigma}_g^{-1}\hat{\bm{\beta}}_g^{bc}$.
\begin{lemma}
\label{lemma:ell}
Suppose assumptions \ref{assumption:OLS} and \ref{assumption:Set1} hold. Then
\begin{align}
\label{equation:Asymell}
n^{1/2}\left( \hat{\bm{\ell}}_g - \bm{\ell}_g \right) \stackrel{\mathcal{D}}{\to} N_K\left( 0, \sigma_g^2 I_K \right).
\end{align}
\end{lemma}
Note the above asymptotic distribution for $\hat{\bm{\ell}}_g$ is exactly the same as if we had observed $\bm{C}$. The next lemma provides asymptotic results for $\hat{\sigma}_g^2$ and $\hat{\rho}$.
\begin{lemma}
\label{lemma:Sigma}
Suppose assumptions \ref{assumption:OLS} and \ref{assumption:Set1} hold. Then
\begin{align}
\label{equation:AsymSigma}
\hat{\sigma}_g^2 = \sigma_g^2 + o_P(1)
\end{align}
and
\begin{align}
\label{equation:Asymrho}
\hat{\rho} = \frac{1}{p}\Tr\left( \hat{\bm{\Sigma}} \right) = \rho + o_P\left( n^{-1/2} \right).
\end{align}
\end{lemma}
The results in both of these lemmas hold regardless of the strength of the standardized confounding effects, so long as $\lambda_K$ is bounded from below and $\lambda_1$ does not grow faster than linearly with $n$. That is, we understand the asymptotic behavior of $\hat{\bm{\ell}}_g$, $\hat{\sigma}_g^2$ and $\hat{\rho}$ even in the scenario when some of the latent factors have strong and others have weak standardized effects.\\
\indent Lastly, we need to use $\bm{Y}_1$ and $\hat{\bm{L}}$ to estimate $\bm{\Omega}^{(OLS)}$. If $\bm{B} = \bm{0}$, then \eqref{equation:Y1.model} implies a reasonable estimator for $\bm{\Omega}^{(OLS)}$ is \eqref{equation:Omega.biased}, the ordinary least squares estimator using $\hat{\bm{L}}$ as the design matrix and $\bm{Y}_1$ as the response. However, in order to guarantee our estimate $\bm{\Omega}^{(OLS)}$ is accurate when the main effect is non-zero, we need the following assumption about the sparsity of $\bm{B}$: 
\begin{assumption}
\label{assumption:B}
Let $\bm{B}_j$ be the $j^{\text{th}}$ column of $\bm{B}$ and
\begin{align*}
\delta_j = \frac{1}{p}\sum\limits_{g=1}^p \bm{1}\left\lbrace \bm{B}_j[g] \neq 0 \right\rbrace
\end{align*}
be the fraction of non-zero entries in $\bm{B}_j$. Then $\max_{g \in [p]}\abs{\bm{B}_j[g]} \leq c_5$ and $\delta_j = o\left( \frac{\lambda_K}{n^{3/2}} \right)$  for some $c_5 > 0$.
\end{assumption}
We note that this is the same sparsity that is needed to prove theorem 3.3 in \citet{CATE}, which only considers the case when $\lambda_K \asymp n$.\par
\indent Using assumption \ref{assumption:B}, we can then prove the following lemma:
\begin{lemma}
\label{lemma:Omega.bc}
Suppose assumptions \ref{assumption:OLS}, \ref{assumption:Set1} and \ref{assumption:B} hold. Further, assume $\frac{\lambda_1}{\lambda_K} \geq c_6^{-1}$ where $c_6 > 0$, i.e. the eigenvalues of the confounding effect matrix are all on the same order of magnitude. Define the estimated bias-corrected effect relating $\bm{C}$ to $\bm{X}$ to be
\begin{align}
\label{equation:Omega.bc}
\hat{\bm{\Omega}}^{(OLS)}_{bc} = \hat{\bm{\Omega}}^{(OLS)} \text{diag}\left( \frac{\hat{\lambda}_1}{\hat{\lambda}_1 - \hat{\rho}}, \ldots, \frac{\hat{\lambda}_K}{\hat{\lambda}_K - \hat{\rho}} \right)
\end{align}
where the naive estimator $\hat{\bm{\Omega}}^{(OLS)}$ is given by \eqref{equation:Omega.biased}. Then,
\begin{equation*}
n^{1/2}\left( \hat{\bm{\Omega}}^{(OLS)}_{bc} - \bm{\Omega}^{(OLS)} \right) = o_P(1)
\end{equation*}
\end{lemma}
We can now state Theorem \ref{theorem:Thm1}:
\begin{theorem}
\label{theorem:Thm1}
Suppose the assumptions of lemma \ref{lemma:Omega.bc} hold and we estimate $\bm{\beta}_g$ as
\begin{align}
\label{equation:beta.bc}
\hat{\bm{\beta}}_g^{bc} = \bm{y}_{1_g} - \hat{\bm{\Omega}}^{(OLS)}_{bc} \hat{\bm{\ell}}_g.
\end{align}
Then the asymptotic distribution for $\hat{\bm{\beta}}_g^{bc}$ is the same as if we had observed the confounding variables $\bm{C}$:
\begin{align}
\label{equation:Dist.beta1}
\frac{n^{1/2}}{\hat{\sigma}_g}\left( \hat{\bm{\beta}}_g^{bc} - \bm{\beta}_g \right) \sim N_d\left( 0, \left( n^{-1}\bm{X}^T\bm{X} \right)^{-1} + \bm{\Omega}^{(OLS)} \bm{\Omega}^{(OLS)T} \right) + o_P(1).
\end{align}
\end{theorem}
Just as we argued in the beginning of section \ref{Properties}, the estimated bias correction term
\begin{align*}
\text{diag}\left( \hat{\lambda}_1/\left( \hat{\lambda}_1 - \hat{\rho} \right), \ldots, \hat{\lambda}_K/\left( \hat{\lambda}_K - \hat{\rho} \right)\right)
\end{align*}
is negligible when $\lambda_K$ is larger than $n^{1/2}$. However, we will show through simulation in section \ref{subsection:Simulations} that ignoring it when the data are not informative for some of the factors discredits inference.\\
\indent An interesting point of discussion is the requirement that the eigenvalues $\lambda_k$ must be of the same order. In real experimental data, it almost always the case the data are only informative for some latent factors and not informative for others, which would manifest itself in some of the $\lambda_k$'s being large and others being small. We therefore extend Lemma \ref{lemma:Omega.bc} and Theorem \ref{theorem:Thm1} in Theorem \ref{thms:Thm2} to relax the assumption that the $\lambda_k$'s be the same order of magnitude.
\begin{theorem}
\label{thms:Thm2}
Suppose assumptions \ref{assumption:OLS}, \ref{assumption:Set1} and \ref{assumption:B} hold, where $\frac{n-d}{p}\bm{L}^T \bm{L} = \text{diag}\left( \lambda_1, \ldots, \lambda_K \right) $. If $\abs{\frac{n-d}{p}\bm{L}_{\cdot r}^T \bm{\Sigma}\bm{L}_{\cdot s}} \leq c_7 \lambda_{\max\left( r,s \right)}$ for some $c_7 > 0$, then for $\hat{\bm{\Omega}}^{(OLS)}_{bc}$ and $\hat{\bm{\beta}}_g^{bc}$ defined in \eqref{equation:Omega.bc} and \eqref{equation:beta.bc}, then
\begin{align}
\label{equation:AsymOmega.bc2}
n^{1/2}\left( \hat{\bm{\Omega}}^{(OLS)}_{bc} - \bm{\Omega}^{(OLS)} \right) = o_P(1)
\end{align}
and
\begin{equation*}
\frac{n^{1/2}}{\hat{\sigma}_g}\left( \hat{\bm{\beta}}_g^{bc} - \bm{\beta}_g \right) \sim N_d\left( 0, \left( n^{-1}\bm{X}^T\bm{X} \right)^{-1} + \bm{\Omega}^{(OLS)} \bm{\Omega}^{(OLS)T} \right) + o_P(1),
\end{equation*}
where $\bm{L}_{\cdot r}$ is the $r^{\text{th}}$ column of $\bm{L}$.
\end{theorem}
The condition on the off-diagonal elements of $\frac{n-d}{p}\bm{L}^T \bm{\Sigma}\bm{L}$ is necessary because we are using truncated SVD to estimate $\bm{L}$, which put into a model based framework is akin to assuming $\bm{\Sigma}$ is a constant multiple of the identity and using the maximum likelihood estimate from a standard Gaussian likelihood as the estimate for $\bm{L}$. When $\bm{\Sigma} \neq \sigma^2 I_p$, we can still use this likelihood model with the correct mean but incorrect variance with the additional minor assumption.\par 
\indent Not only does Theorem \ref{thms:Thm2} allow us to do inference on $\bm{\beta}_g$, but we can use equation \eqref{equation:AsymOmega.bc2} to generalize Theorem 3.5 in \citet{CATE} to do inference on $\bar{\bm{\Omega}}$ when the data are only informative for some of the latent factors and not informative for others:
\begin{corollary}
\label{corollary:OmegaInf}
Suppose the residual matrix $\bar{\bm{\Xi}} \in \mathbb{R}^{n \times K}$ (see Assumption \ref{assumption:OLS}) is independent of $\bm{X}$ and has independent and identically distributed rows $\bm{\xi}_i \in \mathbb{R}^K$ with
\begin{align*}
&\E\left( \bm{\xi}_i \right) = \bm{0}\\
&\E\left( \bm{\xi}_i \bm{\xi}_i^T \right) = \bar{\bm{\Psi}}.
\end{align*}
Suppose further the conditions of Theorem \ref{thms:Thm2} hold and the entries of $\bm{X}$ are bounded from above and below. If the null hypothesis $\bar{\bm{\Omega}} = \bm{0}$ is true, then
\begin{align}
\left( \bm{X}^T \bm{X} \right)^{1/2} \hat{\bm{\Omega}}_{bc}^{(OLS)} \hat{\bm{\Omega}}_{bc}^{(OLS)T} \left( \bm{X}^T \bm{X} \right)^{1/2} \stackrel{\mathcal{D}}{\to} \mathcal{W}_d\left( I_d, K \right),
\end{align}
where $\mathcal{W}_d\left( I_d, K \right)$ is the standard Wishart distribution in $d$ dimensions with $K$ degrees of freedom. If $d = 1$, this is just a $\chi^2_K$ random variable.
\end{corollary}
The proof uses the standardization equations given in \eqref{equations:Standardize} and is a straightforward exercise in multivariate regression analysis once we have proven \eqref{equation:AsymOmega.bc2} in Theorem \ref{thms:Thm2}. Corollary \ref{corollary:OmegaInf} allows us to check if any of the latent factors are significantly correlated with the covariate of interest and can be very useful when trying to uncover the origin of the hidden covariates, as we illustrate with real data in section \ref{subsection:RealData}.

\section{Simulations and data analysis}
\label{section:DataAnalysis}
\subsection{Simulation study}
\label{subsection:Simulations}
In this section we use simulations to illustrate the superior performance of our bias-corrected estimator compared to the uncorrected estimator given in CATE, the software that implements the method proposed in \citet{CATE}. In all of our simulations, we set $n = 100, p = 10^5$ and $K = 10$ to mimic DNA methylation data where $p$ ranges from $3 \times 10^4$ to $8 \times 10^5$, although our results are nearly identical for $p$'s on the order of gene expression data ($p\approx 10^4$). We assumed $d = 1$ and assigned 50 samples to the treatment group and the rest to the control group. We then set $\lambda_1 = n/5, \lambda_K = 1$ and
\begin{align*}
\lambda_k = \left(\frac{n}{5}\right)^{\left( K-k \right)/\left( K-1\right)}.
\end{align*}
For some value of $\bar{\bm{\Omega}} \in \mathbb{R}^{1 \times K}$, we simulated $\bm{B},\bm{L}, \bm{C}, \bm{\Sigma}$ and $\bm{E}$ according to
\begin{align*}
\bm{B}_g &\sim 0.95 \delta_0 + 0.05 N\left( 0, 0.4^2 \right)\\
\bar{\bm{L}}_{gk} &\sim \text{$\pi_k \delta_0 + \left( 1-\pi_k\right)N\left(0, 0.5^2  \right)$ and $\pi_k$ chosen so $\E\left( \bar{\bm{L}}_{\cdot k}^T \bar{\bm{L}}_{\cdot k} \right) = \lambda_k$}\\
\bar{\bm{C}} &\sim MN_{n \times K}\left( \bm{X}\bar{\bm{\Omega}}, I_n, I_K \right)\\
\sigma_g^2 &\sim \text{$\text{Gamma}\left(1/0.5^2, 1/0.5^2 \right)$, i.e. $\E \sigma_g^2 = 1$ and $\text{Var}\left( \sigma_g^2 \right) = 0.5^2$}\\
\bm{E}_{gk} &\sim \frac{\sigma_g}{\sqrt{2}} t_4
\end{align*}
where $t_4$ is the t-distribution with 4 degrees of freedom. Although our theory from section \ref{Properties} assumes the residuals are normally distributed, we simulated data with heavier tails to better mimic real data. We then set $\bar{\bm{\Omega}}$ so that when $\bar{\bm{\Omega}}$ loads exclusively and uniformly on the last $K/2$ columns of $\bar{\bm{L}}$ (i.e. when $\bar{\bm{\Omega}} = \left( \bm{0}_{K/2}^T, \omega \bm{1}_{K/2}^T \right)$), the indirect effect $\bm{X}\bar{\bm{\Omega}} \bar{\bm{\ell}}_g$ contributed approximately 20\% of the variance due to $\bm{X}$ for units $g$ with non-zero direct effect $\beta_g$. That is, we set $\norm{\bar{\bm{\Omega}}}_2^2$ such that
\begin{align*}
0.2 = \frac{\bar{\bm{\Omega}} \E \left( \bar{\bm{\ell}}_g \bar{\bm{\ell}}_g^T\right) \bar{\bm{\Omega}}^T}{\E\left( \beta_g^2 \mid \beta_g \neq 0 \right) + \bar{\bm{\Omega}} \E \left( \bar{\bm{\ell}}_g \bar{\bm{\ell}}_g^T\right) \bar{\bm{\Omega}}^T} = \frac{\norm{\bar{\bm{\Omega}}}_2^2 \sum\limits_{k=K/2+1}^K \lambda_k/(n-2) }{0.4^2 + \norm{\bar{\bm{\Omega}}}_2^2 \sum\limits_{k=K/2+1}^K \lambda_k/(n-2)}
\end{align*}
and let $\bar{\bm{\Omega}}$ take one of two values: 
\begin{align}
\bar{\bm{\Omega}}_1 &= \left( \omega \bm{1}_{K/2}^T, \bm{0}_{K/2}^T \right)\\
\bar{\bm{\Omega}}_2 &= \left( \bm{0}_{K/2}^T , \omega \bm{1}_{K/2}^T \right) \text{ where $\omega = \sqrt{\norm{\bar{\bm{\Omega}}}_2^2 / \left( K/2\right)}$}.
\end{align}
When $\bar{\bm{\Omega}} = \bar{\bm{\Omega}}_1$, the largest components of $\bm{\Omega}^{(OLS)}$ should relatively easy to estimate, since they correspond to the latent factors that are easily estimable from the data (i.e. the factors with the largest $\lambda_k$'s). However, when $\bar{\bm{\Omega}} = \bar{\bm{\Omega}}_2$, proposition \ref{proposition:BiasedOmega} states that uncorrected estimators like the one used in CATE should severely underestimate $\bm{\Omega}^{(OLS)}$, which would lead to greater type I error.\par
\begin{figure}
\centering
\includegraphics[scale=0.3]{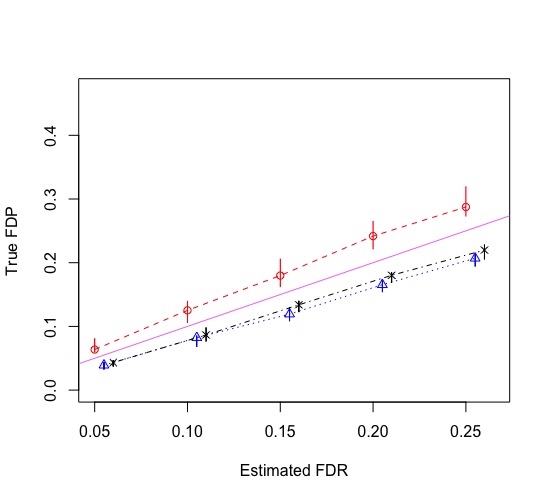}
\includegraphics[scale=0.3]{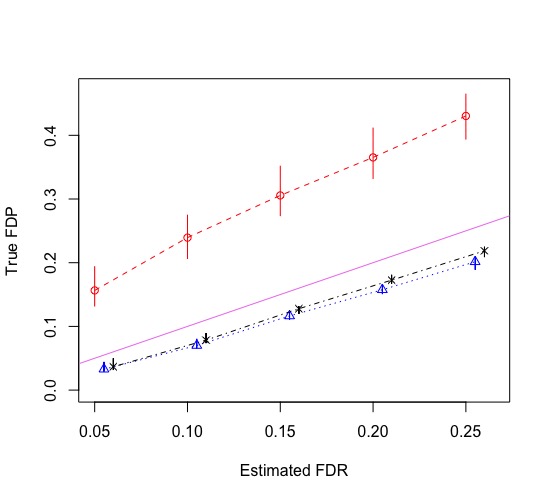}
\caption{Simulation results for $\bar{\bm{\Omega}} = \bar{\bm{\Omega}}_1$ (left) and $\bar{\bm{\Omega}} = \bar{\bm{\Omega}}_2$ (right) when we estimate $\beta_g$ using CATE (red, dashed line with circular points), our bias-corrected method (blue, dotted line with triangular points) or ordinary least squares when $\bm{C}$ is known (black, dot-dashed line with ``x" points), followed by q-value to estimate the false discovery rate (FDR). The points are the median true false discovery proportion (FDP) over all 20 simulations, and the bars are the first and third quartiles. The solid violet line is the $45\degree$ line that passes through the origin.}\label{Fig:SimulationKknown}
\end{figure}

\indent Figure~\ref{Fig:SimulationKknown} provides the estimation results for 40 simulated datasets (20 using $\bar{\bm{\Omega}}_1$ and 20 using $\bar{\bm{\Omega}}_2$) when we estimate the main effect $\beta_g$ using CATE, our bias-corrected estimator \eqref{equation:beta.bc}, and ordinary least squares when $\bm{C}$ is known. We observed that the \textit{P} values reported by CATE were typically biased low even when $\bm{\Omega}^{(OLS)}$ was small, and found that we could perform better inference by performing ordinary least squares with their estimated $\bm{C}$ and comparing the resulting t-statistics to a t-distribution with $n-d-K$ degrees of freedom. We therefore compared the t-statistics from all three methods to a $t_{n-d-K}$ to compute \textit{P} values and judged the performance of each method by comparing the true false discovery proportion (FDP) with the estimated false discovery rate (FDR), estimated with q-value \citep{qvalue}, because this is the inference method and software popular among biologists. Just as one would expect, inference with CATE had large type I error in both simulation scenarios, but especially when latent factors that are correlated with the design matrix are difficult to estimate from the observed data (right panel of Figure~\ref{Fig:SimulationKknown}). However, inference with our bias-corrected estimator was just as accurate as inference with ordinary least squares when $\bm{C}$ was known, even though our simulated data had heavy tails. These results did not change when we over-specified $K$ to be 11 or 12 instead of 10.

\subsection{Data application}
\label{subsection:RealData}
In order to demonstrate the importance of using our bias-corrected estimator and how uncorrected estimators can bias test statistics, we applied our method to re-analyze data from \citet{Jessie}, which studied the correlation between adult asthma and DNA methylation in lung epithelial cells. The authors collected endobronchial brushings from 74 adult patients with a current doctor's diagnosis of asthma and 41 healthy adults and quantified their DNA methylation on $p = 327,271$ methylation sites (i.e. CpGs) using the Infinium Human Methylation 450K Bead Chip \citep{450KChip}. The authors then used ordinary least squares to regress the methylation at each of the $p$ sites onto the mean model subspace that included asthma status, age, ethnicity (European American, African American or Other), gender and smoking status to estimate the effect due to asthma, $\bm{B} \in \mathbb{R}^{p \times 1}$. They found 40,892 CpGs that were differentially methylated between asthmatics and non-asthmatics at a nominal FDR of 5\% (estimated with q-value).\par
\begin{figure}
\centering
\includegraphics[scale=0.3]{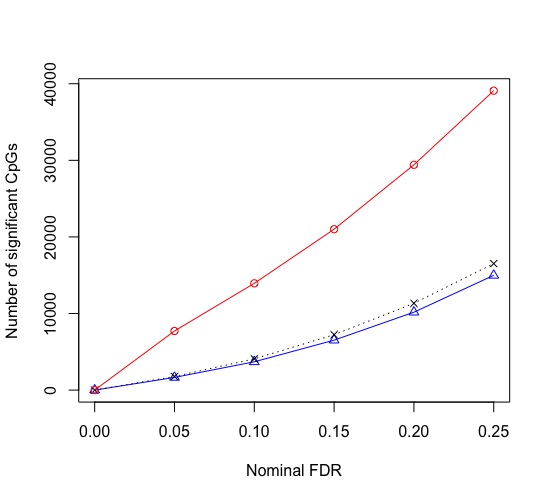}
\caption{A comparison of the number of cytosines (CpGs) whose methylation levels differed between asthmatic subjects and healthy controls at nominal FDR threshold in \citet{Jessie}, using the results from CATE (red, solid line with circular points), our bias-corrected estimator with $K=4$ (blue, solid line with triangular points) and with $K=5$ (black, dotted line with ``x" points) as input into q-value.}\label{Fig:LungResults}
\end{figure}

\indent We then investigated whether or not the strong association between DNA methylation and asthma status was in part due to the fact that lung cell composition may differ between asthmatics and non-asthmatics, with asthmatic patients generally having a greater proportion of airway goblet cells that excrete mucus \citep{GobletAsthma,AsthmaIncreasesGoblet}. We used the same mean model and software provided by \citet{bcv} to estimate that there were an additional $K=4$ latent factors. We then used CATE with the same t-distribution inference modification used in our simulation study and our bias-corrected method to estimate $\bm{\Omega}^{(OLS)}$ and do inference on the effect due to asthma. As observed in Figure \ref{Fig:LungResults}, the results from CATE seem to indicate that asthma has a strong direct effect on DNA methylation, whereas our method implies a mediated signal. In fact, our method identified only 3,600 CpG sites whose methylation levels were correlated with asthma status, whereas CATE confidently identified nearly 14,000 sites at a nominal false discovery rate of 10\%. We then used the results from Corollary \ref{corollary:OmegaInf} and found the \textit{P} value for the null hypothesis that there was no correlation between asthma status and the latent factors to be $5 \times 10^{-10}$, indicating that not only were CATE's estimates for the effect of asthma (while holding all else constant) on methylation likely severely biased, but that cell composition is presumably driving the most of the observed correlation in \citet{Jessie} and the re-analysis with CATE.\par 
\indent To further corroborate the latter, we fit a topic model with $r=7$ topics on the same individuals' gene expression data, which has been shown to cluster bulk RNA-seq samples by tissue and cell type \citep{CountClust_Kushal,CountClust_Taddy}. We then used the $n$-dimensional factor whose corresponding loading was the largest on the \textit{MUC5AC} gene as a proxy for the proportion of goblet cells in each sample, as \textit{MUC5AC} is a unique identifier for goblet cells \citep{MUC5ACGoblet1,MUC5ACGoblet2}. Just as one would expect, asthmatic subjects tended to have a higher estimated proportion of goblet cells (logistic regression p-value = $8 \times 10^{-4}$), which confirmed that the asthmatics in this study tended to have more goblet cells than healthy controls. These results provided additional evidence that our bias-corrected estimator was accounting for cellular heterogeneity, which changes the interpretation as to the source of the observed correlation between asthma status and DNA methylation in \citet{Jessie}.

\section{Discussion}
\label{Discussion}
\indent We have shown that when the data are not informative for the unobserved covariates, previously established weak convergence results do not hold, which can be detrimental to inference. We then provided a bias-corrected estimator for the effects of interest and proved its asymptotic distribution is the same as the ordinary least squares estimator when $\bm{C}$ is observed. Throughout the paper, we assumed $K$ to be known, which is often not the case in real data. However, we found in our data application and simulation results that estimates for $\hat{\bm{\beta}}_g^{bc}$ were not sensitive to over-specifying $K$, which suggests there is a range of $K$'s for which we can perform reliable inference. We also made the critical assumption that $K$ was fixed as $n,p \to \infty$, which is typically not the case in practice. For example, the number of batches in ``omic'' experiments tends to increase as sample size increases, since technicians and machines can only process a fixed number of samples at once. We believe this to be an interesting area of future research.\par
\indent An important assumption we required to guarantee the weak convergence of $\hat{\bm{\beta}}_g^{bc}$ was that the fraction of non-zero entries of $\bm{B}$ needed to be $o\left( \lambda_K/n^{3/2} \right)$, which is the same sparsity assumed in \citet{CATE} and a weaker condition than what is assumed in \citet{Fan1}. If it is safe to assume the entries of $\bm{B}$ are symmetric about zero, generated independently of one-another and are independent of $\bm{L}$, then one can handle stronger signals and replace the sparsity criterion with $o\left( \lambda_K/n \right)$. However, this is still small when the data are not informative for the latent factors. This observation is important for practitioners to be aware of when they are deciding whether or not to include a nuisance covariate (see equation \eqref{equation:NuisanceCov}) in their model or just account for it using our method or some other approach. If one suspects the variable of interest, $\bm{X}$, influences a large fraction of methylation sites or genes, then it would be wise to include the nuisance covariate in the model to avoid incorrectly attributing variability in $\bm{Y}$ due to $\bm{X}$ as coming from $\bm{C}$. If, however, the observed nuisance covariate is a noisy estimate of the actual nuisance variable or if there is no prior belief the factor should affect the response, we recommend correcting for it using the observed data $\bm{Y}$. This is sometimes the case in DNA methylation data when practitioners measure cell composition via flow cytometry or from a noisy DNA methylation reference set measured on ``pure" cell types \citep{HousemanRef,Gervin}.

\section*{Acknowledgement}
We thank Carole Ober and Michelle Stein for comments that have substantially improved this manuscript. The research is supported in part by NIH grants R01-HL129735 and R01-MH101820.

\bibliography{paper-ref.bib}

\setcounter{equation}{0}

\renewcommand{\thefigure}{S\arabic{figure}}
\renewcommand{\theequation}{S\arabic{equation}}
\renewcommand{\thesection}{S\arabic{section}}

\section*{Supplementary material}
\label{section:supplement}
\subsection*{Proofs of all propositions, lemmas, theorems and corollaries}
\label{Proofs}
Recall from equations \eqref{equation:Y1.model}, \eqref{equation:Y2.model} and \eqref{equations:Standardize} that
\begin{align*}
\bm{Y}_1 &= \bm{B} + \bm{L}\bm{\Omega}^{(OLS) T} + \bm{E}_1\\
\bm{Y}_2 &= \bm{L}\bm{C}_2^T + \bm{E}_2
\end{align*}
where $\bm{E}_1 \sim MN_{p \times d}\left( \bm{0}, \bm{\Sigma}, \left( \bm{X}^T \bm{X} \right)^{-1} \right)$ and $\bm{E}_2 \sim MN_{p \times (n-d)} \left( \bm{0}, \bm{\Sigma}, I_{n-d} \right)$ are independent and $(n-d)^{-1} \bm{C}_2^T \bm{C}_2 = I_K$. The estimates for $\bm{L}$ and $\sigma_g^2$ (see equations \eqref{equation:Lhat} and \eqref{equation:C2hat}) were the first $K$ left singular vectors of $\left( n-d \right)^{-1/2} \bm{Y}_2$ multiplied by their corresponding singular values and $(n-d-K)^{-1}\bm{y}_{g_2}^T P_{\hat{C}_2}^{\perp} \bm{y}_{g_2}$, respectively, where $\hat{\bm{C}}_2 = \bm{Y}_2 \hat{\bm{L}}\left( \hat{\bm{L}}^T \hat{\bm{L}} \right)^{-1}$ is our estimate for $\bm{C}_2$. The first goal is to understand the asymptotic properties of $\hat{\bm{L}}$ and $\hat{\bm{C}}_2$, which are essential to all of the proofs that follow.\par
\indent We start by stating and proving Lemmas \ref{lemma:dk_vk} and \ref{lemma:Asyvk.hat} and use their results to prove \eqref{equation:ell.BiasedOmega} from Proposition \ref{proposition:BiasedOmega}, Lemma \ref{lemma:ell} and Lemma \ref{lemma:Sigma} from the main text. For ease of notation, we assume for the statements and proofs of these results that
\begin{equation}
\label{equation:newY.model}
\bm{Y}_{p \times n} = \bm{L}_{p \times K}\bm{C}_{K \times n}^T + \bm{E}_{p \times n}, \quad \bm{E} \sim MN_{p \times n}\left( \bm{0}, \bm{\Sigma}, I_n \right)
\end{equation} 
where $n^{-1} \bm{C}^T \bm{C} = I_K$. We also define
\begin{align}
\label{equation:newC}
\tilde{\bm{C}} &= n^{-1/2}\bm{C}\\
\label{equation:newL}
\tilde{\bm{L}} &= \sqrt{\frac{n}{p}}\bm{L}.
\end{align}
We will lastly define a matrix $\bm{Q} \in \mathbb{R}^{n \times n-K}$ such that $\bm{Q}^T \bm{Q} = I_{n-K}$ and $\bm{Q}_T \tilde{\bm{C}} = \bm{0}_{(n-K)\times K}$. We use a technique developed in \citet{Supp_Debashis} to define the rotated matrix $\bm{F}_{n \times n}$ to be
\begin{align}
\bm{F} &= \left( \begin{matrix}
\tilde{\bm{C}}^T\\
\bm{Q}^T
\end{matrix} \right) \frac{1}{p}\bm{Y}^T \bm{Y} \left( \begin{matrix}
\tilde{\bm{C}} & \bm{Q}
\end{matrix} \right)\nonumber\\
\label{equation:DebashisF}
&= \left[ \begin{matrix}
\left( \tilde{\bm{L}} + \frac{1}{\sqrt{p}}\tilde{\bm{E}}_1 \right)^T \left( \tilde{\bm{L}} + \frac{1}{\sqrt{p}}\tilde{\bm{E}}_1 \right) & \left( \tilde{\bm{L}} + \frac{1}{\sqrt{p}}\tilde{\bm{E}}_1 \right)^T \frac{1}{\sqrt{p}}\tilde{\bm{E}}_2\\
\frac{1}{\sqrt{p}}\tilde{\bm{E}}_2^T \left( \tilde{\bm{L}} + \frac{1}{\sqrt{p}}\tilde{\bm{E}}_1 \right) & \frac{1}{p}\tilde{\bm{E}}_2^T\tilde{\bm{E}}_2
\end{matrix} \right]
\end{align}
where $\tilde{\bm{E}}_1 = \bm{E}\tilde{\bm{C}}$ and $\tilde{\bm{E}}_2 = \bm{E}\bm{Q}$ are independent. Since $\left( \begin{matrix}
\tilde{\bm{C}} & \bm{Q}
\end{matrix} \right)$ is a unitary matrix, the eigenvalues of $\bm{F}$ are also the eigenvalues of $\frac{1}{p}\bm{Y}^T \bm{Y}$. For the remainder of the section, we assume $\begin{pmatrix}
\hat{\bm{V}}_{K \times K}\\
\hat{\bm{Z}}_{(n-K) \times K}
\end{pmatrix}$ are the first $K$ eigenvectors of $\bm{F}$, meaning $\tilde{\bm{C}} \hat{\bm{V}} + \bm{Q}\hat{\bm{Z}}$ are the first $K$ eigenvectors of $\frac{1}{p}\bm{Y}^T \bm{Y}$. Further, since $\tilde{\bm{E}}_1$ and $\tilde{\bm{E}}_2$ are independent, the upper left block of $\bm{F}$ is independent of $\tilde{\bm{E}}_2$. We exploit this by first studying the eigenstructure of the upper left block in Lemma \ref{lemma:dk_vk}, and then using those results to enumerate the asymptotic properties of the first $K$ eigenvalues and eigenvectors of $\bm{F}$ in Lemma \ref{lemma:Asyvk.hat}.

\begin{lemma}
\label{lemma:dk_vk}
Let $\tilde{\bm{L}} \in \mathbb{R}^{p \times K}$, $\tilde{\bm{E}}_1 \sim MN_{p \times K}\left( \bm{0}, \bm{\Sigma}, I_K \right)$ and $\tilde{\bm{N}} = \tilde{\bm{L}} + \frac{1}{\sqrt{p}}\tilde{\bm{E}}_1$. Assume $\tilde{\bm{L}}^T \tilde{\bm{L}} = \text{diag}\left( \lambda_1, \ldots, \lambda_K \right)$ where the $\lambda_k$'s are the same as those given in assumption \ref{assumption:Set1} (with $d = 0$) and $\bm{\Sigma}$ follows assumption \ref{assumption:OLS}. If , $d_k^2 = \bm{\lambda}_k\left( \tilde{\bm{N}}^T \tilde{\bm{N}} \right)$ and $\bm{v}_k$ are the $k^{\text{th}}$ eigenvalue and eigenvector of $\tilde{\bm{N}}^T \tilde{\bm{N}}$, then
\begin{align}
\label{equation:Asydk}
\frac{d_k^2}{\lambda_k} = 1 + \frac{\rho}{\lambda_k} + \mathcal{O}_P\left( \frac{1}{\sqrt{\lambda_k p}} \right)
\end{align}
and
\begin{align}
\label{equation:Asyvk}
\bm{v}_k =& \left( 1 + \mathcal{O}_P\left( \frac{1}{\lambda_k p} \right) \right)\bm{e}_k + \mathcal{O}_P\left( \frac{1}{\sqrt{\lambda_1 p}} \right)\bm{e}_1 + \cdots + \mathcal{O}_P\left( \frac{1}{\sqrt{\lambda_{k-1} p}} \right)\bm{e}_{k-1}\\ 
&+ \mathcal{O}_P\left( \frac{1}{\sqrt{\lambda_k p}} \right)\bm{e}_{k+1} + \cdots + \mathcal{O}_P\left( \frac{1}{\sqrt{\lambda_k p}} \right)\bm{e}_{K}\nonumber
\end{align}
where $\bm{e}_k$ are the standard basis vectors in $\mathbb{R}^{K}$.
\end{lemma}

\begin{proof}
First, $\tilde{\bm{N}}^T \tilde{\bm{N}} = \tilde{\bm{L}}^T \tilde{\bm{L}} + \rho I_K + \frac{1}{\sqrt{p}}\tilde{\bm{L}}^T \tilde{\bm{E}}_1 + \frac{1}{\sqrt{p}}\tilde{\bm{E}}_1^T \tilde{\bm{L}} + \bm{B}$ where the entries of $\bm{B}$ are $\mathcal{O}_P\left( \frac{1}{\sqrt{p}} \right)$. Let $\bm{R}\bm{R}^T = \tilde{\bm{L}}^T\bm{\Sigma}\tilde{\bm{L}}$ where $\bm{R}$ is a lower triangular matrix. By Cauchy-Schwartz, we have that $\bm{R}_k^T = \bm{R}[k,] = \mathcal{O}\left( \sqrt{\lambda_k} \right)$. We also note that $\frac{1}{\sqrt{p}}\tilde{\bm{L}}^T\tilde{\bm{E}}_1 \sim \bm{R} \bm{M}$ where the entries of $\bm{M} \in \mathbb{R}^{K \times K}$ are $\mathcal{O}_P\left( \frac{1}{\sqrt{p}} \right)$. If we let the columns of $\bm{M}$ be $\bm{M}_s$, then $\left[ \bm{R} \bm{M} \right]_{ks} = \bm{R}_k^T \bm{M}_s = \mathcal{O}_P\left( \frac{\sqrt{\lambda_k}}{\sqrt{p}} \right)$. Next, define the matrix $\bm{A}^{(1)} \in \mathbb{R}^{K \times K}$ to be
\begin{align*}
\bm{A}^{(1)} &= \frac{1}{\lambda_1}\tilde{\bm{N}}^T \tilde{\bm{N}} = \left( \begin{matrix}
\mu_1 & a_{12} & \cdots & a_{1K}\\
a_{21} & \mu_2 & \cdots & a_{2K}\\
\vdots & \vdots & \ddots & \vdots\\
a_{K1} & a_{K2} & \cdots & \mu_K
\end{matrix} \right)
\end{align*}
where 
\begin{align*}
\mu_k &= \frac{\lambda_k + \rho}{\lambda_1} + \frac{2}{\lambda_1} \bm{R}_k^T\bm{M}_k + \frac{1}{\lambda_1} \bm{B}_{kk}\\
a_{ks} &= \frac{1}{\lambda_1} \bm{R}_k^T \bm{M}_s + \frac{1}{\lambda_1} \bm{R}_s^T \bm{M}_k + \frac{1}{\lambda_1}\bm{B}_{sk} = \mathcal{O}_P\left( \frac{\sqrt{\lambda_k}}{\lambda_1\sqrt{p}} \right)
\end{align*}
for $k < s$. Our goal is to break $\bm{A}^{(1)}$ into $K$ rank one pieces, each of which are approximately orthogonal. The procedure is as follows:
\begin{enumerate}
\item Define $\bm{A}_1 = \bm{A}[,1], \bm{A}_2 = \left( 0, \bm{A}[2,2:K] \right)^T, \ldots, \bm{A}_K = \left( \underbrace{0, \ldots, 0}_{\text{$K-1$ 0's}}, \bm{A}[K,K] \right)^T$.
\item We wish to first modify $\bm{A}_1$ and $\bm{A}_2$ so that they are orthogonal. To do this, we will add $\epsilon_2$ to $\bm{A}_2[1]$ and remove $\epsilon_2$ from $\bm{A}_1[2]$. That is, we define $\bm{A}_{1_2} = \bm{A}_1 + \epsilon_2 \bm{e}_2$ and $\bm{A}_{2_2} = \bm{A}_2 - \epsilon_2 \bm{e}_1$ such that
\begin{equation*}
0 = \bm{A}_{1_2}^T \bm{A}_{2_2} = \bm{A}_1^T\bm{A}_2 + \epsilon_2 \mu_2 - \epsilon_2 \mu_1 = a_{12}\mu_2 + \epsilon_2 \mu_2 - \epsilon_2 \mu_1 + \mathcal{O}_P\left( \frac{\sqrt{\lambda_2}}{\lambda_1^{3/2} p} \right)
\end{equation*}
meaning $\epsilon_2 = \frac{a_{12}\mu_2}{\mu_1 - \mu_2} + \mathcal{O}_P\left( \frac{\sqrt{\lambda_2}}{\lambda_1^{3/2} p} \right) = \mathcal{O}_P\left( \frac{\lambda_2}{\lambda_1^{3/2}\sqrt{p}} \right)$. We now have $\bm{A}_{1_2}^T \bm{A}_{2_2} = 0$.
\item Define $\bm{A}_{1_k} = \bm{A}_{1_{k-1}} + \epsilon_k \bm{e}_k$ and $\bm{A}_{k_2} = \bm{A}_{k} - \epsilon_k \bm{e}_1$ inductively:
\begin{equation*}
0 = \left( \bm{A}_{1_{k-1}} + \epsilon_k \bm{e}_k \right)^T \left( \bm{A}_k - \epsilon_k \bm{e}_1 \right) = \bm{A}_{1_{k-1}}^T\bm{A}_k + \epsilon_k \mu_k - \epsilon_k \mu_1 = a_{1k}\mu_k + \epsilon_k \mu_k - \epsilon_k \mu_1 + \mathcal{O}_P\left( \frac{\sqrt{\lambda_k}}{\lambda_1^{3/2} p} \right)
\end{equation*}
meaning $\epsilon_{k} = \frac{a_{1k}\mu_k}{\mu_1 - \mu_k} + \mathcal{O}_P\left( \frac{\sqrt{\lambda_k}}{\lambda_1^{3/2} p} \right) = \mathcal{O}_P\left( \frac{\lambda_k}{\lambda_1^{3/2}\sqrt{p}} \right)$. 
\item After we complete this process $K-1$ times to get $\bm{A}_{1_K}$, we now have for $s < K$
\begin{align*}
\bm{A}_{1_K}^T \bm{A}_{s_2} &= \left( \bm{A}_1 + \epsilon_2 \bm{e}_2 + \cdots + \epsilon_K \bm{e}_K \right)^T \left( \bm{A}_s - \epsilon_s \bm{e}_1 \right) = \left( \bm{A}_1 + \epsilon_2 \bm{e}_2 + \cdots + \epsilon_s \bm{e}_s \right)^T \left( \bm{A}_s - \epsilon_s \bm{e}_1 \right) + \\
&\left( \epsilon_{s+1}\bm{e}_{s+1} + \cdots + \epsilon_{K}\bm{e}_K \right)^T \left( \bm{A}_s - \epsilon_s \bm{e}_1 \right) = 0 + \epsilon_{s+1}a_{s,s+1} + \cdots \epsilon_{K}a_{s,K} \\
&= \mathcal{O}_P\left( \frac{\lambda_{s+1}}{\lambda_1^{3/2} \sqrt{p}} \frac{\sqrt{\lambda}_s}{\lambda_1 \sqrt{p}} \right) \leq \mathcal{O}_P\left( \frac{\lambda_s^{3/2}}{\lambda_1^{3/2}} \frac{1}{\lambda_1 p} \right)
\end{align*}
and $\bm{A}_{1_K}^T \bm{A}_{1_K} = \mu_1^2 + \mathcal{O}_P\left( \frac{1}{\lambda_1 p} \right)$, meaning $\norm{\bm{A}_{1_K}}_2 = \mu_1 + \mathcal{O}_P\left( \frac{1}{\lambda_1 p} \right)$.
\end{enumerate}

We now have
\begin{align*}
\bm{A}^{(1)} &= \underbrace{\left( \begin{matrix}
\bm{A}_{1_K} & \rightarrow\\
\downarrow & \bm{0}_{(K-1) \times (K-1)}
\end{matrix} \right)}_{\bm{B}^{(1)}} + \underbrace{\left( \begin{matrix}
0 & \uparrow & \bm{0}_{1 \times (K-2)}\\
\leftarrow & \bm{A}_{2_2} & \rightarrow\\
\bm{0}_{(K-2) \times 1} & \downarrow & \bm{0}_{(K-2) \times (K-2)}
\end{matrix} \right)}_{\bm{B}^{(2)}} + \cdots + \underbrace{\left( \begin{matrix}
\bm{0}_{(K-1) \times (K-1)} & \uparrow\\
\leftarrow & \bm{A}_{K_2}
\end{matrix} \right)}_{\bm{B}^{(K)}} \\
&= \begin{pmatrix}
\mu_1 & a_{12} + \epsilon_2 & \cdots & a_{1K} + \epsilon_K\\
a_{12} + \epsilon_2 & 0 & \cdots & 0\\
\vdots & \vdots & \ddots & \vdots\\
a_{1K} + \epsilon_K & 0 & \cdots & 0
\end{pmatrix} + \begin{pmatrix}
0 & - \epsilon_2 & 0 & \cdots & 0\\
- \epsilon_2 & \mu_2 & a_{23} & \cdots & a_{2K}\\
\vdots & \vdots & \vdots & \cdots & \vdots\\
0 & a_{2K} & 0 & \cdots & 0
\end{pmatrix} + \cdots + \begin{pmatrix}
0 & \cdots & 0 & -\epsilon_K\\
\vdots & \cdots & \vdots & \vdots\\
0 & \cdots & 0 & 0\\
-\epsilon_K & \cdots & 0 & \mu_K
\end{pmatrix}
\end{align*}

Define $\bm{u}_{1_K} = \frac{1}{\norm{\bm{A}_{1_K}}_2}\bm{A}_{1_K} = \left( 1, \frac{a_{12} + \epsilon_2}{\mu_1}, \ldots, \frac{a_{1K} + \epsilon_K}{\mu_1} \right)^T + \mathcal{O}_P\left( \frac{1}{\lambda_1 p} \right)$. Then $\bm{B}^{(1)} = \mu_1 \bm{u}_{1_K} \bm{u}_{1_K}^T + \mathcal{O}_P\left( \frac{1}{\lambda_1 p} \right)$. Further,
\begin{equation*}
\norm{\bm{B}^{(s)}\bm{u}_{1_K}}_2 = \norm{\begin{pmatrix}
-\epsilon_s\frac{a_{1s} + \epsilon_s}{\norm{\bm{A}_{1_K}}_2}\\
0\\
\vdots\\
0\\
\frac{1}{\norm{\bm{A}_{1_K}}_2} \bm{A}_{s_2}^T \bm{A}_{1_K}\\
a_{s,s+1}\frac{a_{1s} + \epsilon_s}{\norm{\bm{A}_{1_K}}_2}\\
\vdots\\
a_{s,K}\frac{a_{1s} + \epsilon_s}{\norm{\bm{A}_{1_K}}_2}
\end{pmatrix}}_2 \leq \mathcal{O}_P\left( \frac{1}{\lambda_1 p} \right)
\end{equation*}
which means $\bm{A}^{(1)}\bm{u}_{1_K} = \mu_1 \bm{u}_{1_K} + \mathcal{O}_P\left( \frac{1}{\lambda_1 p} \right)$. We define $\delta = \bm{u}_{1_K}^T \bm{A}^{(1)}\bm{u}_{1_K} = \mu_1 + \mathcal{O}_P\left( \frac{1}{\lambda_1 p} \right)$ and $\gamma = \norm{\bm{A}^{(1)}\bm{u}_{1_K} - \delta \bm{u}_{1_K}}_2 = \mathcal{O}_P\left( \frac{1}{\lambda_1 p} \right)$. By Weyl's Thm, the eigenvalues of $\bm{A}^{(1)}$ are $\lambda_k + \mathcal{O}_P\left( \frac{1}{\sqrt{\lambda_1 p}} \right)$, so if $\xi$ is the second largest eigenvalue of $\bm{A}^{(1)}$, $\xi = \mu_2 + \mathcal{O}_P\left( \frac{1}{\sqrt{\lambda_1 p}} \right)$, meaning $f = \delta - \xi = \frac{\lambda_1 - \lambda_2}{\lambda_1} + \mathcal{O}_P\left( \frac{1}{\sqrt{\lambda_1 p}} \right)$. By Thm 3.6 in \citet{Supp_Tang}, we have
\begin{enumerate}
\item $\exists$ an eigenvalue $\lambda_{\gamma}$ of $\bm{A}^{(1)}$ s.t. $\lambda_{\gamma} \in \left[ \delta - \gamma, \delta + \gamma \right]$, i.e. $\lambda_{\gamma} = \mu_1 + \mathcal{O}_P\left( \frac{1}{\lambda_1 p} \right)$
\item If $\bm{v}_{\gamma}$ is the eigenvalue corresponding to $\lambda_{\gamma}$ and $f > \gamma$,
\begin{equation*}
\norm{\bm{v}_{\gamma} - \bm{u}_{1_K}^T\bm{v}_{\gamma} \bm{u}_{1_K} }_2 \leq \frac{2 \gamma}{f - \gamma} = \mathcal{O}_P\left( \frac{1}{\lambda_1 p} \right)
\end{equation*}
Let $G_{\lambda_{\gamma}, n,p} = \left\lbrace \text{$\lambda_{\gamma}$ is the maximum eigenvalue of $\bm{A}^{(1)}$} \right\rbrace$. Then
\begin{equation*}
P\left( \abs{\bm{\lambda}_1 \left( \bm{A}^{(1)}\right) - \mu_1} \geq M \right) \leq P\left( \abs{\lambda_{\gamma} - \mu_1}\geq M, G_{\lambda_{\gamma}, n,p} \right) + P\left( G_{\lambda_{\gamma}, n,p}^c \right) \leq P\left( \abs{\delta - \mu_1} \geq M \right) + P\left( G_{\lambda_{\gamma}, n,p}^c \right)
\end{equation*}
Since $P\left( G_{\lambda_{\gamma}, n,p}^c \right) \to 0$ and $\abs{\lambda_{\gamma}- \mu_1} = \mathcal{O}_P\left( \frac{1}{\lambda_1 p} \right)$, $\frac{d_1^2}{\lambda_1} = \bm{\lambda}_1 \left( \bm{A}^{(1)}\right) = \mu_1 + \mathcal{O}_P\left( \frac{1}{\lambda_1 p} \right)$. We can apply an identical procedure to show that $\norm{\bm{v}_{1} - \bm{u}_{1_K}^T\bm{v}_{1} \bm{u}_{1_K} }_2 = \mathcal{O}_P\left( \frac{1}{\lambda_1 p} \right)$. Since $\bm{v}_1$ and $\bm{u}_{1_K}$ are unit vectors, we must have $\bm{u}_{1_K}^T\bm{v}_{1} = \pm 1 + \mathcal{O}_P\left( \left( \frac{1}{\lambda_1 p} \right)^2 \right)$. That is, we know $\bm{v}_{1}$ up to sign parity.
\end{enumerate}
We then have
\begin{equation*}
\bm{A}^{(2)} = \frac{1}{\lambda_2}\left( \lambda_1\bm{A}^{(1)} - d_1^2 \bm{v}_1 \bm{v}_1^T \right) = \frac{\lambda_1}{\lambda_2}\bm{B}^{(2)} + \cdots + \frac{\lambda_1}{\lambda_2}\bm{B}^{(K)} + \mathcal{O}_P\left( \frac{1}{\lambda_2 p} \right). 
\end{equation*}
Since $\epsilon_k \frac{\lambda_1}{\lambda_2} = \mathcal{O}_P\left( \frac{\lambda_k}{\lambda_2}\frac{1}{\sqrt{\lambda_1 p}} \right)$, all off-diagonal entries of the above matrix at most $\mathcal{O}_P\left( \frac{1}{\sqrt{\lambda_2 p}} \right)$. We can then apply the exact same procedure as we did above to show that $\frac{d_k^2}{\lambda_k} = \lambda_k + \frac{\rho}{\lambda_k} + \mathcal{O}_P\left( \frac{1}{\sqrt{\lambda_k p}} \right)$ and $\bm{v}_k = \begin{pmatrix}
\mathcal{O}_P\left( \frac{1}{\sqrt{\lambda_k p}} \right)\\
\vdots\\
1 + \mathcal{O}_P\left( \frac{1}{\lambda_1 p} \right)\\
\vdots\\
\mathcal{O}_P\left( \frac{1}{\sqrt{\lambda_k p}} \right)
\end{pmatrix}$ (this part is omitted). Lastly, for $s < k$,
\begin{equation*}
0 = \bm{v}_s^T \bm{v}_k = \bm{v}_k[s]\bm{v}_s[s] + \mathcal{O}_P\left( \frac{1}{\lambda_k p} \right) + \bm{v}_s[k] \bm{v}_k[k] = \bm{v}_k[s] + \mathcal{O}_P\left( \frac{1}{\sqrt{\lambda_s p}} \right)
\end{equation*}
meaning $\bm{v}_k[s] = \mathcal{O}_P\left( \frac{1}{\sqrt{\lambda_s p}} \right)$.

\end{proof}

\begin{center}
\line(1,0){425}
\end{center}

We use $\tilde{\bm{E}}_1$, $\tilde{\bm{E}}_2$, $\tilde{\bm{N}}$, $d_k$ and $\bm{v}_k$ defined in Lemma \ref{lemma:dk_vk} in the remainder of the paper. We also define
\begin{align}
\label{equation:RandomR}
\bm{R} = \frac{1}{p}\tilde{\bm{E}}_2^T \tilde{\bm{E}}_2 - \rho I_{n-K}
\end{align}
and let $\bm{V} = \begin{bmatrix}
\bm{v}_1 & \cdots & \bm{v}_K
\end{bmatrix} $, $\bm{U} = \begin{bmatrix}
\bm{u}_1 \cdots \bm{u}_K
\end{bmatrix}$ be the first $K$ right and left singular values of $\tilde{\bm{N}}$. By Theorem 5.39 in \citet{Supp_Vershynin}, $\norm{\bm{R}}_2 = \mathcal{O}_P\left( \sqrt{\frac{n}{p}} \right)$. The next lemma uses what we have established in Lemma \ref{lemma:dk_vk} to prove convergence properties of the first $K$ eigenvalues and eigenvectors of $\bm{F}$ (see \eqref{equation:DebashisF}).

\begin{lemma}
\label{lemma:Asyvk.hat}
Suppose the probability model for $\bm{Y}$ is given by \eqref{equation:newY.model} and that assumptions \ref{assumption:OLS} and \ref{assumption:Set1} hold for $d = 0$ ($d$ is the number of columns in $\bm{X}$). Then
\begin{equation}
\label{equation:lambdak.hat}
\hat{\lambda}_k = \bm{\lambda}_k\left( \bm{F} \right) = d_k^2 + \mathcal{O}_P\left( \frac{n}{p} \right).
\end{equation}
Define $\begin{bmatrix}
\hat{\bm{v}}_k\\
\hat{\bm{z}}_k
\end{bmatrix}$, $\hat{\bm{v}}_k \in \mathbb{R}^K$ and $\hat{\bm{z}}_k \in \mathbb{R}^{n-K}$ to be the $k^{\text{th}}$ eigenvector of $\bm{F}$. Then
\begin{equation}
\label{equation:vk.hat}
\hat{\bm{v}}_k = \bm{v}_k + \bm{\epsilon}_k, \quad \norm{\bm{\epsilon}_k}_2 = \mathcal{O}_P\left( \frac{n}{\lambda_k p} \right).
\end{equation}
and 
\begin{equation}
\label{equation:zk.hat}
\hat{\bm{z}}_k = \frac{d_k}{\lambda_k \sqrt{p}}\tilde{\bm{E}}_2^T \bm{u}_k + \bm{R}\frac{d_k}{\lambda_k^2 \sqrt{p}}\tilde{\bm{E}}_2^T \bm{u}_k + \mathcal{O}_P\left( \left( \frac{n}{\lambda_k p} \right)^{3/2} + \frac{n^{1/2}}{p \lambda_k} \right)
\end{equation}
where $d_k$ and $\bm{v}_k$ are given by \eqref{equation:Asydk} and \eqref{equation:Asyvk} and $\bm{u}_k$ is the $k^{\text{th}}$ left singular vector of $\bm{Y} \tilde{\bm{C}}$. Further, if $\frac{n}{p}\bm{L}_{\cdot r}^T \bm{\Sigma}\bm{L}_{\cdot s} \leq c \lambda_{\max\left( r,s \right)}$ then
\begin{equation}
\label{equation:epsilon}
\bm{\epsilon}_k[s] = o_P\left( \frac{\lambda_k}{\lambda_s}n^{-1/2} \right).
\end{equation}
\end{lemma}

\begin{proof}
First, define
\begin{align*}
\bm{F}^{(1)} = \bm{F} = \lambda_1 \begin{bmatrix}
	\bm{\hat{A}}_1 & \bm{H}_1\\
	\bm{H}_1^T & \bm{J}_1 \end{bmatrix}.
\end{align*}
We immediately observe from the expression for $\bm{F}$ in \eqref{equation:DebashisF} that $\frac{\hat{\lambda}_1}{\lambda_1} = \frac{d_1^2}{\lambda_1} + \mathcal{O}_P\left( \sqrt{\frac{n}{\lambda_1 p}} \right) = \frac{\lambda_1 + \rho}{\lambda_1} + \mathcal{O}_P\left( \sqrt{\frac{n}{\lambda_1 p}} \right)$ by Weyl's Theorem. The eigenvalue equations for $\bm{F}^{(1)}$ are
	\begin{equation*}
	\frac{\hat{\lambda}_1}{\lambda_1}\hat{\bm{v}}_1 = \hat{\bm{A}}_1 \hat{\bm{v}}_1 + \bm{H}_1 \hat{\bm{z}}_1
	\end{equation*}
	\begin{equation*}
	\frac{\hat{\lambda}_1}{\lambda_1}\hat{\bm{z}}_1 = \bm{H}_1^T \hat{\bm{v}}_1 + \bm{J}_1 \hat{\bm{z}}_1
	\end{equation*}
	\begin{equation}
	\Rightarrow \hat{\bm{z}}_1 = \left( \frac{\hat{\lambda}_1}{\lambda_1}I_{n-K} - \bm{J}_1 \right)^{-1}\bm{H}_1^T \hat{\bm{v}}_1, \quad \frac{\hat{\lambda}_1}{\lambda_1}\hat{\bm{v}}_1 = \hat{\bm{A}}_1 \hat{\bm{v}}_1 + \bm{H}_1 \left( \frac{\hat{\lambda}_1}{\lambda_1}I_{n-K} - \bm{J}_1 \right)^{-1}\bm{H}_1^T \hat{\bm{v}}_1
	\end{equation}
where $\bm{H}_1 = \frac{1}{\lambda_1} \left( \tilde{\bm{L}} + \frac{1}{\sqrt{p}}\tilde{\bm{E}}_1 \right)^T \frac{1}{\sqrt{p}}\tilde{\bm{E}}_2$ and $\frac{\hat{\lambda}_1}{\lambda_1}I_{n-K} - \bm{J}_1 = \frac{\hat{\lambda}_1 - \rho}{\lambda_1}I_{n-K} - \frac{1}{\lambda_1}R$ is invertable with high probability, since $\frac{\hat{\lambda}_1}{\lambda_1} = \frac{\lambda_1 + \rho}{\lambda_1} + \mathcal{O}_P\left( \sqrt{\frac{n}{\lambda_1 p}} \right)$ and $R = \frac{1}{p}\tilde{\bm{E}}_2^T\tilde{\bm{E}}_2  - \rho I_{n-K} = \mathcal{O}_P\left( \sqrt{\frac{n}{p}} \right)$. Therefore, $\norm{\left( \frac{\hat{\lambda}_1}{\lambda_1}I_{n-K} - \bm{J}_1 \right)^{-1}}_2 = \mathcal{O}_P\left( 1 \right)$ and $\norm{\bm{H}_1}_2 = \mathcal{O}_P\left( \frac{n^{1/2}}{\sqrt{\lambda_1 p}} \right)$, meaning $\norm{\bm{H}_1 \left( \frac{\hat{\lambda}_1}{\lambda_1}I_{n-K} - \bm{J}_1 \right)^{-1}\bm{H}_1^T}_2 = \mathcal{O}_P\left( \frac{n}{\lambda_1 p} \right)$. Since $\hat{\bm{A}}_1 = \bm{A}^{(1)}$ (see Lemma \ref{lemma:dk_vk}),
\begin{equation*}
\frac{\hat{\lambda}_1}{\lambda_1} = \bm{\lambda}_1\left( \bm{A}^{(1)} \right) + \mathcal{O}_P\left( \frac{n}{\lambda_1 p} \right) = \frac{d_1^2}{\lambda_1} + \mathcal{O}_P\left( \frac{n}{\lambda_1 p} \right)
\end{equation*}
by Weyl's Theorem. To determine the behavior of $\hat{\bm{v}}_1$, we first notice that since $\hat{\bm{z}}_1^T \hat{\bm{z}}_1 = \mathcal{O}_P\left( \frac{n}{\lambda_1 p} \right)$ and $\norm{\hat{\bm{v}}_1}_2^2 + \norm{\hat{\bm{z}}_1}_2^2 = 1$, $\norm{\hat{\bm{v}}_1}_2 = 1 + \mathcal{O}_P\left( \frac{n}{\lambda_1 p} \right)$. Further, because $\bm{H}_1 \left( \frac{\hat{\lambda}_1}{\lambda_1}I_{n-K} - \bm{J}_1 \right)^{-1}\bm{H}_1^T = \mathcal{O}_P\left( \frac{n}{\lambda_1 p} \right)$,
\begin{equation*}
\hat{\bm{v}}_1 = \frac{1}{\sqrt{\norm{\bm{v}_1 + \mathcal{O}_P\left( \frac{n}{\lambda_1 p} \right)}_2^2 + \norm{\hat{\bm{z}}_1}_2^2}} \left( \bm{v}_1 + \mathcal{O}_P\left( \frac{n}{\lambda_1 p} \right) \right) = \bm{v}_1 + \mathcal{O}_P\left( \frac{n}{\lambda_1 p} \right).
\end{equation*}
Using these above relations, we can get an expression for $\hat{\bm{z}}_1$:
\begin{align*}
\hat{\bm{z}}_1 &= \left( \frac{\hat{\lambda}_1}{\lambda_1}I_{n-K} - \bm{J}_1 \right)^{-1}\bm{H}_1^T \hat{\bm{v}}_1 = \underbrace{\frac{1}{\lambda_1}\left( \frac{\hat{\lambda}_1}{\lambda_1}I_{n-K} - \frac{1}{\lambda_1 p}\tilde{\bm{E}}_2^T \tilde{\bm{E}}_2 \right)^{-1} \frac{1}{\sqrt{p}}\tilde{\bm{E}}_2^T \left( \tilde{\bm{L}} + \frac{1}{\sqrt{p}}\tilde{\bm{E}}_1 \right)}_{\mathcal{O}_P\left( \sqrt{\frac{n}{\lambda_1 p}} \right)} \underbrace{\hat{\bm{v}}_1}_{\bm{v}_1 + \mathcal{O}_P\left( \frac{n}{\lambda_1 p} \right)} \\
 &=\frac{1}{\lambda_1}\left( \underbrace{\frac{\hat{\lambda}_1 - \rho}{\lambda_1}}_{1 + \mathcal{O}_P\left( \frac{1}{\sqrt{\lambda_1 p}} + \frac{n}{\lambda_1 p} \right)}I_{n-K} - \frac{1}{\lambda_1}\bm{R} \right)^{-1}\frac{1}{\sqrt{p}}\tilde{\bm{E}}_2^T \underbrace{\left( \tilde{\bm{L}} + \frac{1}{\sqrt{p}}\tilde{\bm{E}}_1 \right)}_{\bm{U}\bm{D}\bm{V}^T} \bm{v}_1 + \mathcal{O}_P\left( \left( \frac{n}{\lambda_1 p} \right)^{3/2} \right)\\
 & = \frac{d_1}{\lambda_1 \sqrt{p}}\left( I_{n-K} - \frac{1}{\lambda_1}\bm{R} \right)^{-1} \tilde{\bm{E}}_2^T \bm{u}_1 + \mathcal{O}_P\left( \left( \frac{n}{\lambda_1 p} \right)^{3/2} + \frac{n^{1/2}}{p \lambda_1} \right)\\
 & = \frac{d_1}{\lambda_1 \sqrt{p}} \tilde{\bm{E}}_2^T \bm{u}_1 + \frac{d_1}{\lambda_1^2 \sqrt{p}}\bm{R}\tilde{\bm{E}}_2^T \bm{u}_1 + \mathcal{O}_P\left( \left( \frac{n}{\lambda_1 p} \right)^{3/2} + \frac{n^{1/2}}{p \lambda_1} \right)
\end{align*}
since
\begin{align*}
\norm{\left( I_{n-K} - \frac{1}{\lambda_1}\bm{R} \right)^{-1} - \left( I_{n-K} + \frac{1}{\lambda_1}\bm{R} \right)}_2 = \mathcal{O}\left( \norm{\frac{1}{\lambda_1^2}\bm{R}^2}_2 \right) = \mathcal{O}_P\left( \frac{n}{\lambda_1^2 p} \right).
\end{align*}
We can then determine $\frac{\hat{\lambda}_k}{\lambda_k}$, $\hat{\bm{v}}_k$ and $\hat{\bm{z}}_k$ by induction:\\
\\
\indent First, we assume
\begin{enumerate}
\item $\hat{\lambda}_k = d_k^2 + \mathcal{O}_P\left( \frac{n}{p} \right)$, $\hat{\bm{v}}_k = \bm{v}_k + \mathcal{O}_P\left( \frac{n}{\lambda_k p} \right)$
\item \begin{equation*}
\hat{\bm{z}}_k = \frac{d_k}{\lambda_k \sqrt{p}}\tilde{\bm{E}}_2^T \bm{u}_k + \bm{R}\frac{d_k}{\lambda_k^2 \sqrt{p}}\tilde{\bm{E}}_2^T \bm{u}_k + \mathcal{O}_P\left( \left( \frac{n}{\lambda_k p} \right)^{3/2} + \frac{n^{1/2}}{p \lambda_k} \right)
\end{equation*}
\item \begin{align*}
\lambda_{k}\bm{H}_k^T &= \frac{1}{\sqrt{p}}\tilde{\bm{E}}_2^T \tilde{\bm{N}} - \hat{\lambda}_1\hat{\bm{z}}_1\hat{\bm{v}}_1^T - \cdots - \hat{\lambda}_{k-1}\hat{\bm{z}}_{k-1}\hat{\bm{v}}_{k-1}^T\\
& = \mathcal{O}_P\left( \frac{1}{\sqrt{\lambda_1}}\sqrt{\frac{n}{p}} \right)\bm{v}_1^T + \cdots + \mathcal{O}_P\left( \frac{1}{\sqrt{\lambda_{k-1}}}\sqrt{\frac{n}{p}} \right)\bm{v}_{k-1}^T + \frac{1}{\sqrt{p}}\tilde{\bm{E}}_2^T \sum	\limits_{\ell = k}^K d_{\ell}\bm{u}_{\ell}\bm{v}_{\ell}^T\\
& + \mathcal{O}_P\left( \frac{1}{\sqrt{\lambda_{k-1}}}\left( \frac{n}{p} \right)^{3/2} + \frac{n^{1/2}}{p} \right)
\end{align*}
\end{enumerate}

Then
\begin{enumerate}

\item \begin{align*}
\lambda_{k+1}\bm{H}_{k+1}^T &= \frac{1}{\sqrt{p}}\tilde{\bm{E}}_2^T \tilde{\bm{N}} - \hat{\lambda}_1\hat{\bm{z}}_1\hat{\bm{v}}_1^T - \cdots - \hat{\lambda}_{k}\hat{\bm{z}}_{k}\hat{\bm{v}}_{k}^T\\
 &= \mathcal{O}_P\left( \frac{1}{\sqrt{\lambda_1}}\sqrt{\frac{n}{p}} \right)\bm{v}_1^T + \cdots + \mathcal{O}_P\left( \frac{1}{\sqrt{\lambda_{k-1}}}\sqrt{\frac{n}{p}} \right)\bm{v}_{k-1}^T +  \frac{d_k}{\sqrt{p}}\tilde{\bm{E}}_2^T \bm{u}_{k}\bm{v}_{k}^T\\
 & + \frac{1}{\sqrt{p}}\tilde{\bm{E}}_2^T \sum	\limits_{\ell = k+1}^K d_{\ell}\bm{u}_{\ell}\bm{v}_{\ell}^T + \mathcal{O}_P\left( \frac{1}{\sqrt{\lambda_{k-1}}}\left( \frac{n}{p} \right)^{3/2} + \frac{n^{1/2}}{p} \right)\\
 & - \underbrace{\left( \frac{\hat{\lambda}_k}{\lambda_k} \right)}_{1 + \frac{\rho}{\lambda_k} + \mathcal{O}_P\left( \frac{1}{\sqrt{\lambda_k p}} + \frac{n}{p \lambda_k} \right)} \underbrace{\frac{d_k}{\sqrt{p}}\tilde{\bm{E}}_2^T\bm{u}_k\hat{\bm{v}}_k^T}_{\frac{d_k}{\sqrt{p}}\tilde{\bm{E}}_2^T\bm{u}_k \bm{v}_k^T + \mathcal{O}_P\left( \frac{1}{\sqrt{\lambda_k}}\left( \frac{n}{p} \right)^{3/2} \right)}\\
 & - \underbrace{\left( \frac{\hat{\lambda}_k}{\lambda_k} \right)\bm{R} \frac{d_k}{\lambda_k \sqrt{p}}\tilde{\bm{E}}_2^T\bm{u}_k\hat{\bm{v}}_k^T}_{\mathcal{O}_P\left( \frac{n}{p \sqrt{\lambda_k}} \right)\bm{v}_k^T + \mathcal{O}_P\left( \frac{1}{\sqrt{\lambda_k}}\left( \frac{n}{p} \right)^{3/2} \right)}\\
 & = \mathcal{O}_P\left( \frac{1}{\sqrt{\lambda_1}}\sqrt{\frac{n}{p}} \right)\bm{v}_1^T + \cdots + \mathcal{O}_P\left( \frac{1}{\sqrt{\lambda_{k}}}\sqrt{\frac{n}{p}} \right)\bm{v}_{k}^T + \mathcal{O}_P\left( \frac{1}{\sqrt{\lambda_k}}\left( \frac{n}{p} \right)^{3/2} +  \frac{n^{1/2}}{p} \right)\\
 & + \frac{1}{\sqrt{p}}\tilde{\bm{E}}_2^T \sum\limits_{\ell = k+1}^K d_{\ell}\bm{u}_{\ell}\bm{v}_{\ell}^T
\end{align*}
This means $\norm{\bm{H}_{k+1}}_2 = \mathcal{O}_P\left( \sqrt{\frac{n}{\lambda_{k+1} p}} \right)$.

\item \begin{align*}
\lambda_{k+1}\hat{\bm{A}}_{k+1} &= \tilde{\bm{N}}^T \tilde{\bm{N}} - \hat{\lambda}_1 \hat{\bm{v}}_1 \hat{\bm{v}}_1^T - \cdots - \hat{\lambda}_k \hat{\bm{v}}_k \hat{\bm{v}}_k^T = \tilde{\bm{N}}^T \tilde{\bm{N}} - d_1^2 \bm{v}_1 \bm{v}_1^T - \cdots - d_k^2 \bm{v}_k \bm{v}_k^T + \mathcal{O}_P\left( \frac{n}{p} \right)\\
& = \bm{A}^{(k+1)} + \mathcal{O}_P\left( \frac{n}{p} \right)
\end{align*}

\item \begin{equation*}
\lambda_{k+1}\bm{J}_{k+1} = \frac{1}{p}\tilde{\bm{E}}_2^T \tilde{\bm{E}}_2 - \hat{\lambda}_1 \hat{\bm{z}}_1\hat{\bm{z}}_1^T - \cdots - \hat{\lambda}_k \hat{\bm{z}}_k \hat{\bm{z}}_k^T = \frac{1}{p}\tilde{\bm{E}}_2^T \tilde{\bm{E}}_2 + \mathcal{O}_P\left( \frac{n}{p} \right)
\end{equation*}
By the above expressions for $\hat{\bm{A}}_{k+1}$, $\bm{H}_{k+1}$ and $\bm{J}_{k+1}$, $\frac{\hat{\lambda}_{k+1} - \rho}{\lambda_{k+1}} = \frac{d_{k+1}^2 - \rho}{\lambda_{k+1}} + \mathcal{O}_P\left( \sqrt{\frac{n}{\lambda_{k+1} p}} \right) = 1 + \mathcal{O}_P\left( \sqrt{\frac{n}{\lambda_{k+1} p}} \right)$. Therefore, $\left( \frac{\hat{\lambda}_{k+1}}{\lambda_{k+1}}I_{n-K} - \bm{J}_{k+1} \right) = \left( \frac{\hat{\lambda}_{k+1} - \rho}{\lambda_{k+1}}I_{n-K} - \frac{1}{\lambda_{k+1}}\bm{R} + \mathcal{O}_P\left( \frac{n}{\lambda_{k+1}p} \right) \right)$ is invertible with high probability. We then compute the eigenvalue equations to get:

\item 
	\begin{enumerate}
	\item \begin{align*}
&\frac{\hat{\lambda}_{k+1}}{\lambda_{k+1}}\hat{\bm{v}}_{k+1} = \bm{A}^{(k+1)}\hat{\bm{v}}_{k+1}\\
 & + \frac{1}{\lambda_{k+1}^2 p}\left(\sum\limits_{\ell = k+1}^K d_{\ell} \bm{v}_{\ell} \bm{u}_{\ell}^T \right)\tilde{\bm{E}}_2 \left( \frac{\hat{\lambda}_{k+1} - \rho}{\lambda_{k+1}}I_{n-K} - \frac{1}{\lambda_{k+1}}\bm{R} + \mathcal{O}_P\left( \frac{n}{\lambda_{k+1}p} \right) \right)^{-1} \tilde{\bm{E}}_2^T \left(\sum\limits_{\ell = k+1}^K d_{\ell}\bm{u}_{\ell}\bm{v}_{\ell}^T\right) \hat{\bm{v}}_{k+1}\\
 &+\mathcal{O}_P\left( \frac{n}{\lambda_{k+1}p} \right)\hat{\bm{v}}_{k+1} = \bm{A}^{(k+1)}\hat{\bm{v}}_{k+1} + \mathcal{O}_P\left( \frac{n}{\lambda_{k+1}p} \right)\hat{\bm{v}}_{k+1}
\end{align*}

	\item \begin{align*}
	\hat{\bm{z}}_{k+1} &= \left( \frac{\hat{\lambda}_{k+1} - \rho}{\lambda_{k+1}}I_{n-K} - \frac{1}{\lambda_{k+1}}\bm{R} + \mathcal{O}_P\left( \frac{n}{\lambda_{k+1} p} \right) \right)^{-1} \frac{1}{\lambda_{k+1}\sqrt{p}}\tilde{\bm{E}}_2^T \sum\limits_{\ell = k+1}^K d_{\ell}\bm{u}_{\ell}\bm{v}_{\ell}^T \hat{\bm{v}}_{k+1}\\
	&+\mathcal{O}_P\left( \frac{1}{ \lambda_{k+1}\sqrt{\lambda_1}}\sqrt{\frac{n}{p}} \right)\bm{v}_1^T \hat{\bm{v}}_{k+1} + \cdots + \mathcal{O}_P\left( \frac{1}{ \lambda_{k+1}\sqrt{\lambda_{k}}}\sqrt{\frac{n}{p}} \right)\bm{v}_{k}^T \hat{\bm{v}}_{k+1}\\
	& + \mathcal{O}_P\left( \frac{1}{\sqrt{\lambda_k} \lambda_{k+1}}\left( \frac{n}{p} \right)^{3/2} \right) + \mathcal{O}_P\left( \frac{n^{1/2}}{ \lambda_{k+1} p} \right)
	\end{align*}
	Therefore, $\norm{\hat{\bm{z}}_{k+1}}_2 = \mathcal{O}_P\left( \sqrt{\frac{n}{\lambda_{k+1} p}} \right)$, meaning $\norm{\hat{\bm{v}}_{k+1}}_2 = 1 - \mathcal{O}_P\left( \frac{n}{\lambda_{k+1} p} \right)$. From what we have in part a), $\hat{\bm{v}}_{k+1} = \bm{v}_{k+1} + \mathcal{O}_P\left( \frac{n}{\lambda_{k+1}p} \right)$ and $\hat{\lambda}_{k+1} = d^2_{k+1} + \mathcal{O}_P\left( \frac{n}{p} \right)$. We can then modify our expression for $\hat{\bm{z}}_{k+1}$ to get
	\item \begin{align*}
	\hat{\bm{z}}_{k+1} &= \left( \underbrace{\frac{\hat{\lambda}_{k+1} - \rho}{\lambda_{k+1}}}_{1 + \mathcal{O}_P\left( \frac{1}{\sqrt{\lambda_{k+1} p}} + \frac{n}{p \lambda_{k+1}} \right)} I_{n-K} - \frac{1}{\lambda_{k+1}}\bm{R} \right)^{-1} \frac{1}{\lambda_{k+1}\sqrt{p}}\tilde{\bm{E}}_2^T \sum\limits_{\ell = k+1}^K d_{\ell}\bm{u}_{\ell}\bm{v}_{\ell}^T \hat{\bm{v}}_{k+1}\\
	&+\mathcal{O}_P\left( \frac{1}{ \lambda_{k+1}^2\sqrt{\lambda_1}}\left( \frac{n}{p} \right)^{3/2} \right) + \cdots + \mathcal{O}_P\left( \frac{1}{ \lambda_{k+1}^2\sqrt{\lambda_k}}\left( \frac{n}{p} \right)^{3/2} \right)\\
    &+ \mathcal{O}_P\left(\left( \frac{n}{p \lambda_{k+1}} \right)^{3/2} \right) + \mathcal{O}_P\left( \frac{n^{1/2}}{ \lambda_{k+1} p} \right)\\
	&=\frac{1}{\lambda_{k+1}\sqrt{p}}\tilde{\bm{E}}_2^T \sum\limits_{\ell = k+1}^K d_{\ell}\bm{u}_{\ell}\bm{v}_{\ell}^T \hat{\bm{v}}_{k+1} + \frac{1}{\lambda_{k+1}^2\sqrt{p}}\bm{R}\tilde{\bm{E}}_2^T \sum\limits_{\ell = k+1}^K d_{\ell}\bm{u}_{\ell}\bm{v}_{\ell}^T \hat{\bm{v}}_{k+1} + \mathcal{O}_P\left(\left( \frac{n}{p \lambda_{k+1}} \right)^{3/2} \right)\\
	 &+ \mathcal{O}_P\left( \frac{n^{1/2}}{ \lambda_{k+1} p} \right)\\
	&=\frac{d_{k+1}}{\lambda_{k+1}\sqrt{p}}\tilde{\bm{E}}_2^T \bm{u}_{k+1} + \frac{d_{k+1}}{\lambda_{k+1}^2\sqrt{p}}\bm{R}\tilde{\bm{E}}_2^T\bm{u}_{k+1} + \mathcal{O}_P\left(\left( \frac{n}{p \lambda_{k+1}} \right)^{3/2} \right) + \mathcal{O}_P\left( \frac{n^{1/2}}{ \lambda_{k+1} p} \right)
	\end{align*}
	\end{enumerate}

\end{enumerate}

This proves \eqref{equation:lambdak.hat}, \eqref{equation:vk.hat} and \eqref{equation:zk.hat}. It remains to show \eqref{equation:epsilon}.\\
\indent Since $\bm{F}$ is symmetric with distinct eigenvalues (wp1), for $s < k$ (i.e. $\lambda_s > \lambda_k$),
\begin{equation*}
		0 = \hat{\bm{v}}_{s}^T \hat{\bm{v}}_{k} + \hat{\bm{z}}_{s}^T \hat{\bm{z}}_{k} = \left( \bm{v}_s + \bm{\epsilon}_s \right)^T \left( \bm{v}_k + \bm{\epsilon}_k \right) + \hat{\bm{z}}_{s}^T \hat{\bm{z}}_{k} = 0 + \underbrace{\bm{\epsilon}_s^T \hat{\bm{v}}_{k}}_{\mathcal{O}_P\left( \frac{n}{p \lambda_s} \right)} + \underbrace{\bm{v}_s^T \bm{\epsilon}_k}_{\bm{\epsilon}_k[s] + \mathcal{O}_P\left( \frac{1}{\sqrt{\lambda_s p}}\frac{n}{p \lambda_k} + \frac{1}{p\lambda_s}\frac{n}{p \lambda_k} \right)} + \hat{\bm{z}}_{s}^T \hat{\bm{z}}_{k}.
\end{equation*}
By assumption, $\mathcal{O}_P\left( \frac{1}{\sqrt{\lambda_s p}}\frac{n}{p \lambda_k} + \frac{1}{p\lambda_s}\frac{n}{p \lambda_k} \right) = o_P\left( \frac{\lambda_k}{\lambda_s}n^{-1/2} \right)$. Therefore, if we can show 
$\hat{\bm{z}}_{s}^T \hat{\bm{z}}_{k} = o_P\left( \frac{\lambda_k}{\lambda_s}n^{-1/2} \right)$, we must have $\bm{\epsilon}_k[s] =  o_P\left( \frac{\lambda_k}{\lambda_s}n^{-1/2} \right)$. By our above expression for $\hat{\bm{z}}_k$,
\begin{align*}
\hat{\bm{z}}_s^T \hat{\bm{z}}_k &= \left( \frac{d_s}{\lambda_s\sqrt{p}}\tilde{\bm{E}}_2^T \bm{u}_s + \frac{d_s}{\lambda_s^2\sqrt{p}}\bm{R}\tilde{\bm{E}}_2^T\bm{u}_s + \mathcal{O}_P\left(\left( \frac{n}{p \lambda_{s}} \right)^{3/2} +  \frac{n^{1/2}}{ \lambda_{s} p} \right) \right)^T \left(\frac{d_{k}}{\lambda_{k}\sqrt{p}}\tilde{\bm{E}}_2^T \bm{u}_{k} \right.\\
&\left. + \frac{d_{k}}{\lambda_{k}^2\sqrt{p}}\bm{R}\tilde{\bm{E}}_2^T\bm{u}_{k} + \mathcal{O}_P\left(\left( \frac{n}{p \lambda_{k}} \right)^{3/2} +  \frac{n^{1/2}}{ \lambda_{k} p} \right)\right)
\end{align*}

We see that
\begin{enumerate}
\item \begin{equation*}
\mathcal{O}_P\left(\left( \frac{n}{p \lambda_{k}} \right)^{3/2} +  \frac{n^{1/2}}{ \lambda_{k} p} \right) \norm{\hat{\bm{z}}_s}_2 = \mathcal{O}_P\left( \left( \frac{n}{p} \right)^2 \frac{1}{\lambda_k^{3/2} \sqrt{\lambda_s}} + \frac{n}{p \lambda_k}\frac{1}{\sqrt{p \lambda_s}} \right) = o_P\left( \frac{\lambda_k}{\lambda_s}n^{-1/2} \right)
\end{equation*}
\item \begin{equation*}
\norm{\frac{d_s}{\lambda_s^2\sqrt{p}}\bm{R}\tilde{\bm{E}}_2^T\bm{u}_s + \mathcal{O}_P\left(\left( \frac{n}{p \lambda_{s}} \right)^{3/2} +  \frac{n^{1/2}}{ \lambda_{s} p} \right)}_2 \norm{\hat{\bm{z}}_k}_2 \underbrace{\leq}_{\bm{R} = \mathcal{O}_P\left( \sqrt{\frac{n}{p}} \right)} \mathcal{O}_P\left( \frac{n}{p \lambda_s} \sqrt{\frac{n}{p \lambda_k}} \right) = o_P\left( \frac{\lambda_k}{\lambda_s}n^{-1/2} \right)
\end{equation*}
\end{enumerate}
Therefore,
\begin{align*}
\hat{\bm{z}}_s^T \hat{\bm{z}}_k &= \left( \frac{d_s}{\lambda_s\sqrt{p}}\tilde{\bm{E}}_2^T \bm{u}_s \right)^T \left(\frac{d_{k}}{\lambda_{k}\sqrt{p}}\tilde{\bm{E}}_2^T \bm{u}_{k} + \frac{d_{k}}{\lambda_{k}^2\sqrt{p}}\bm{R}\tilde{\bm{E}}_2^T\bm{u}_{k} \right) + o_P\left( \frac{\lambda_k}{\lambda_s}n^{-1/2} \right)\\
& = \frac{d_s d_k}{\lambda_s \lambda_k p} \bm{u}_s^T \tilde{\bm{E}}_2 \tilde{\bm{E}}_2^T \bm{u}_k + \frac{d_s d_k}{\lambda_s \lambda_k^2 p}\bm{u}_s^T \tilde{\bm{E}}_2 \bm{R} \tilde{\bm{E}}_2^T \bm{u}_k + o_P\left( \frac{\lambda_k}{\lambda_s}n^{-1/2} \right).
\end{align*}
Splitting this up term by term, we get
\begin{enumerate}
\item \begin{equation*}
\frac{d_s d_k}{\lambda_s \lambda_k p} \bm{u}_s^T \tilde{\bm{E}}_2 \tilde{\bm{E}}_2^T \bm{u}_k \underbrace{\sim}_{\bm{M} \sim MN_{p \times n \left( 0, I_p, I_n \right)}} \frac{d_s d_k}{\lambda_s \lambda_k p} \bm{u}_s^T\bm{\Sigma}^{1/2} \bm{M}\bm{M}^T \bm{\Sigma}\bm{u}_k
\end{equation*}

Define $\bm{U}_{s,k} = \begin{pmatrix}
\bm{u}_s & \bm{u}_k
\end{pmatrix} $ and $\bm{W} = \left( \bm{U}_{s,k}^T \bm{\Sigma} \bm{U}_{s,k} \right)^{1/2}$. Then
\begin{align*}
\frac{d_s d_k}{\lambda_s \lambda_k p} \bm{u}_s^T \tilde{\bm{E}}_2 \tilde{\bm{E}}_2^T \bm{u}_k &= \left[ \bm{U}_{s,k}^T\tilde{\bm{E}}_2 \tilde{\bm{E}}_2^T \bm{U}_{s,k} \right]_{1,2} \underbrace{\sim}_{\bm{M} \sim MN_{n \times 2}\left( 0, I_n, I_2 \right)} \left[ \frac{d_s d_k n}{\lambda_s \lambda_k p}\bm{W} \left( \frac{1}{n}\bm{M}^T \bm{M} \right)\bm{W} \right]_{1,2}\\
& = \frac{d_s d_k n}{\lambda_s \lambda_k p}\left[ \bm{W}^2 + \mathcal{O}_P\left( \frac{1}{n^{1/2}} \right) \right]_{1,2}\\ 
&= \frac{d_s d_k n}{\lambda_s \lambda_k p} \bm{u}_s^T \bm{\Sigma} \bm{u}_k + \mathcal{O}_P\left( \frac{n^{1/2}}{\sqrt{\lambda_s \lambda_k} p} \right) = \frac{d_s d_k n}{\lambda_s \lambda_k p} \bm{u}_s^T \bm{\Sigma} \bm{u}_k + o_P\left( \frac{\lambda_k}{\lambda_s}n^{-1/2} \right).
\end{align*}

If $\bm{\Sigma} = \sigma^2 I_p$, we would be done. However, if $\bm{\Sigma}$ were arbitrary then under no assumptions $\bm{u}_s^T \bm{\Sigma} \bm{u}_k = \mathcal{O}_P(1)$, meaning $\frac{d_s d_k n}{\lambda_s \lambda_k p} \bm{u}_s^T \bm{\Sigma} \bm{u}_k = \mathcal{O}_P\left( \frac{1}{\sqrt{\lambda_s \lambda_k}}\frac{n}{p} \right)$ which is not necessarily $o_P\left( \frac{\lambda_k}{\lambda_s}n^{-1/2} \right)$. To see this, if $\lambda_s = n$ and $\lambda_k = 1$ then $\mathcal{O}_P\left( \frac{1}{\sqrt{\lambda_s \lambda_k}}\frac{n}{p} \right) = \mathcal{O}_P\left( \frac{n^{1/2}}{p} \right)$, which is not $o_P\left( \frac{\lambda_k}{\lambda_s}n^{-1/2} \right)$. We will use the assumption that $\frac{n}{p}\bm{L}\left[ ,r \right]^T \bm{\Sigma}\bm{L}\left[ ,s \right] \leq c \lambda_{\max\left( r,s \right)}$ in the statement of the lemma to show that $\bm{u}_s^T \bm{\Sigma} \bm{u}_k = \mathcal{O}_P\left( \sqrt{\frac{\lambda_k}{\lambda_s}} \right)$. If this were the case, we would have $\frac{d_s d_k n}{\lambda_s \lambda_k p} \bm{u}_s^T \bm{\Sigma} \bm{u}_k = \mathcal{O}_P\left( \frac{n}{\lambda_s p} \right) = o_P\left( \frac{\lambda_k}{\lambda_s}n^{-1/2} \right)$. Lemma \ref{lemma:UtU} at the end of the section proves $\bm{u}_s^T \bm{\Sigma} \bm{u}_k = \mathcal{O}_P\left( \sqrt{\frac{\lambda_k}{\lambda_s}} \right)$.

\item Recall that $\bm{R} = \frac{1}{p}\tilde{\bm{E}}_2^T \tilde{\bm{E}}_2 - \rho I_{n-K}$. We will prove a lemma that shows $\frac{1}{p}\bm{u}_s^T \tilde{\bm{E}}_2 \bm{R} \tilde{\bm{E}}_2^T \bm{u}_k = \mathcal{O}_P\left( \left( \frac{n}{p} \right)^2 + \frac{n}{p}\frac{1}{\sqrt{p}} \right)$. Once we prove the lemma, we will have $\frac{d_s d_k}{\lambda_s \lambda_k^2 p}\bm{u}_s^T \tilde{\bm{E}}_2 \bm{R} \tilde{\bm{E}}_2^T \bm{u}_k = o_P\left( \frac{\lambda_k}{\lambda_s}n^{-1/2} \right)$. We prove this in lemma \ref{lemma:RandomR} at the end of the section.

\end{enumerate}

This proves \eqref{equation:epsilon} and completes the proof.

\end{proof}

\begin{center}
\line(1,0){425}
\end{center}

Using the results from lemmas \ref{lemma:dk_vk} and \ref{lemma:Asyvk.hat}, we can prove \eqref{equation:ell.BiasedOmega} from Proposition \ref{proposition:BiasedOmega} and lemmas \ref{lemma:ell} and \ref{lemma:Sigma}. In the two proofs below, we assume the data $\bm{Y}$ are distributed as in Lemmas \ref{lemma:dk_vk} and \ref{lemma:Asyvk.hat}

\begin{proof}[of lemma \ref{lemma:ell}]

For the sake of notation I will assume that $\bm{Y} = \bm{L}\bm{C}^T + \bm{E}$, i.e. $\bm{Y}$ follows \eqref{equation:newY.model}. Define $\bm{y}_g$ and $\tilde{\bm{e}}_{i,g}$ to be the $g^{\text{th}}$ row of $\bm{Y}$ and $\tilde{\bm{E}}_i$.
\begin{equation*}
n^{1/2}\hat{\bm{\ell}}_g = \hat{\tilde{\bm{C}}}^T\bm{y}_g = \left( \hat{\bm{V}}^T\tilde{\bm{C}}^T + \hat{\bm{Z}}^T \bm{Q}^T \right)\bm{y}_g = n^{1/2}\hat{\bm{V}}^T\bm{\ell}_g + \hat{\bm{V}}^T \tilde{\bm{e}}_{g,1} + \hat{\bm{Z}}^T \tilde{\bm{e}}_{2,g}
\end{equation*}
\begin{enumerate}
\item \begin{equation*}
n^{1/2}\hat{\bm{V}}^T\bm{\ell}_g = n^{1/2}\bm{\ell}_g + n^{1/2}\mathcal{O}_P\left( \frac{1}{\sqrt{p \lambda_K}} + \frac{n}{p \lambda_K} \right)
\end{equation*}
\item \begin{equation*}
\hat{\bm{V}}^T \tilde{\bm{e}}_{g,1} \sim N\left( 0, \sigma_g^2 I_K \right) + \mathcal{O}_P\left( \frac{1}{\sqrt{p \lambda_K}} + \frac{n}{p \lambda_K} \right)
\end{equation*}
\item \begin{equation*}
\hat{\bm{z}}_k^T \tilde{\bm{e}}_{2,g} = \frac{d_k}{\lambda_k \sqrt{p}}\bm{u}_k[g] \tilde{\bm{e}}_{g,2}^T\tilde{\bm{e}}_{g,2} + \frac{d_k}{\lambda_k \sqrt{p}}\bm{u}_k[-g]^T \tilde{\bm{E}}_2[-g,]\tilde{\bm{e}}_{g,2} + \mathcal{O}_P\left( \frac{n n^{1/2}}{p \lambda_k} \right)
\end{equation*}
where $\bm{u}_k[g] = \frac{1}{d_k}\sqrt{\frac{n}{p}}\left( \bm{\ell}_g + \frac{1}{n^{1/2}}\tilde{\bm{e}}_{g,1} \right)^T \bm{v}_k = \mathcal{O}_P\left( \frac{n^{1/2}}{\sqrt{p} d_k} \right)$. Therefore, $\frac{d_k}{\lambda_k \sqrt{p}}\bm{u}_k[g] \tilde{\bm{e}}_{g,2}^T\tilde{\bm{e}}_{g,2} = \mathcal{O}_P\left( \frac{n^{1/2}n}{p \lambda_k} \right)$. Lastly,
\begin{equation*}
\bm{u}_k[-g]^T \tilde{\bm{E}}_2[-g,]\tilde{\bm{e}}_{g,2} \sim N\left( 0, \bm{u}_k[-g]^T\bm{\Sigma}[-g]\bm{u}_k[-g]\tilde{\bm{e}}_{g,2}^T \tilde{\bm{e}}_{g,2} \right) = \mathcal{O}_P\left( n^{1/2} \right).
\end{equation*}
\end{enumerate}
Therefore, $\hat{\bm{Z}}^T \tilde{\bm{e}}_{2,g} = \mathcal{O}_P\left( \frac{n^{1/2}n}{p \lambda_K} + \sqrt{\frac{n}{p \lambda_K}} \right)$, which means $n^{1/2}\left( \hat{\bm{\ell}}_g - \bm{\ell}_g \right) \stackrel{\mathcal{D}}{\to} N_K\left( 0, \sigma_g^2 I_K \right)$.

\end{proof}

\begin{center}
\line(1,0){425}
\end{center}

\begin{proof}[of lemma \ref{lemma:Sigma}]

Once we estimate $\bm{C}_2$ by SVD, we simply let
\begin{equation*}
\hat{\sigma}_g^2 = \frac{1}{n-d-K}\bm{y}_{2,g}^T P_{\bm{C}_2}^{\perp}\bm{y}_{2,g}
\end{equation*}
for each site $g = 1, \ldots, p$. I will show 2 things:
\begin{itemize}
\item $\hat{\sigma}_g^2 = \sigma_g^2 + \mathcal{O}_P\left( \frac{1}{n^{1/2}} + \sqrt{\frac{n}{\lambda_K p}} \right) = \sigma_g^2 + o_P(1)$.
\item $\hat{\rho} = \frac{1}{p}\Tr\left( \hat{\bm{\Sigma}} \right) = \rho + \mathcal{O}_P\left( \frac{1}{\sqrt{p \lambda_K}} + \frac{n}{\lambda_K p} \right) = \rho + o_P\left( n^{-1/2} \right)$.
\end{itemize}

We define the estimated scaled covariates $\hat{\bm{W}} = \frac{1}{n^{1/2}}\hat{\bm{C}} = \tilde{\bm{C}} \hat{\bm{V}} + \bm{Q} \hat{\bm{Z}} \in \mathbb{R}^{n \times K}$, where $\hat{\bm{V}}$, $\hat{\bm{Z}}$, $\tilde{\bm{C}}^T$ and $\bm{Q}^T$ are given in lemmas \ref{lemma:dk_vk} and \ref{lemma:Asyvk.hat}. Also, define $\bm{\epsilon} = \begin{bmatrix}
\bm{\epsilon}_1 & \cdots & \bm{\epsilon}_K
\end{bmatrix}$, where $\bm{\epsilon}_k$ is as defined in \eqref{equation:vk.hat} of Lemma \ref{lemma:Asyvk.hat}. First,
\begin{align*}
\left( n-K \right)\hat{\sigma}_g^2 &= \bm{y}_g^T \bm{y}_g - \bm{y}_g^T P_{\hat{\bm{W}}}\bm{y}_g = \bm{y}_g^T \bm{y}_g - \bm{y}_g^T \hat{\bm{W}} \hat{\bm{W}}^T\bm{y}_g = \bm{y}_g^T \bm{y}_g - \bm{y}_g^T \left( \tilde{\bm{C}} \hat{\bm{V}} + \bm{Q} \hat{\bm{Z}} \right) \left( \hat{\bm{V}}^T \tilde{\bm{C}}^T + \hat{\bm{Z}}^T \bm{Q}^T \right)\bm{y}_g\\
& =\left( \bm{y}_g^T \bm{y}_g - \bm{y}_g^T \tilde{\bm{C}} \hat{\bm{V}}\hat{\bm{V}}^T \tilde{\bm{C}}^T\bm{y}_g \right) - 2\bm{y}_g^T \tilde{\bm{C}} \hat{\bm{V}}\hat{\bm{Z}}^T \bm{Q}^T\bm{y}_g - \bm{y}_g^T \bm{Q} \hat{\bm{Z}} \hat{\bm{Z}}^T \bm{Q}^T\bm{y}_g
\end{align*}
I will go through the above expression piece by piece to analyze both $\hat{\sigma}_g^2$ and $\hat{\rho}$.
\begin{enumerate}
\item \begin{align*}
\bm{y}_g^T \bm{y}_g - \bm{y}_g^T \tilde{\bm{C}} \hat{\bm{V}}\hat{\bm{V}}^T \tilde{\bm{C}}^T\bm{y}_g& = \bm{y}_g^T \bm{y}_g - \bm{y}_g^T \tilde{\bm{C}} \tilde{\bm{C}}^T\bm{y}_g + 2\bm{y}_g^T \tilde{\bm{C}}\bm{\delta}^T\tilde{\bm{C}}^T\bm{y}_g + \bm{y}_g^T \tilde{\bm{C}}\bm{\delta}^T \bm{\delta}\tilde{\bm{C}}^T \bm{y}_g\\
& = (n-K)\hat{\sigma}_{g,\text{OLS}}^2 + (n-K)\mathcal{O}_P\left( \frac{1}{\sqrt{\lambda_k p}} + \frac{n}{p \lambda_k} \right)
\end{align*}
where $\bm{\delta} = \hat{\bm{V}} - I_K$.
	\begin{enumerate}
	\item \begin{equation*}
	\frac{1}{n-K}\left( \bm{y}_g^T \bm{y}_g - \bm{y}_g^T \tilde{\bm{C}} \hat{\bm{V}}\hat{\bm{V}}^T \tilde{\bm{C}}^T\bm{y}_g \right) = \hat{\sigma}_{g,\text{OLS}}^2 + \mathcal{O}_P\left( \frac{1}{\sqrt{\lambda_k p}} + \frac{n}{p \lambda_k} \right) = \sigma_g^2 + \mathcal{O}_P\left( n^{-1/2} \right)
	\end{equation*}
	
		\item \begin{enumerate}
		\item \begin{equation*}
		\frac{1}{(n-K) p}\sum\limits_{g=1}^p \left( \bm{y}_g^T \bm{y}_g - \bm{y}_g^T \tilde{\bm{C}} \tilde{\bm{C}}^T\bm{y}_g \right) = \frac{1}{p}\sum\limits_{g=1}^p \hat{\sigma}_{g, \text{OLS}}^2 = \rho + \mathcal{O}_P\left( \frac{1}{\sqrt{np}} \right)
	\end{equation*}	
		\item \begin{align*}
	\abs{\frac{1}{np}\sum\limits_{g=1}^p \bm{y}_g^T \tilde{\bm{C}} \bm{\delta}^T\tilde{\bm{C}}^T\bm{y}_g }&=\abs{\frac{1}{p}\sum\limits_{g=1}^p \left( \bm{\ell}_g + \frac{1}{n^{1/2}}\tilde{\bm{e}}_{g,1} \right)^T \bm{\delta}^T \left( \bm{\ell}_g + \frac{1}{n^{1/2}}\tilde{\bm{e}}_{g,1} \right)}\\
	& \leq \underbrace{\left( \frac{1}{p}\sum\limits_{g=1}^p \left( \bm{\ell}_g + \frac{1}{n^{1/2}}\tilde{\bm{e}}_{g,1} \right)^T \bm{\delta}^T \bm{\delta} \left( \bm{\ell}_g + \frac{1}{n^{1/2}}\tilde{\bm{e}}_{g,1} \right) \right)^{1/2}}_{\mathcal{O}_P\left( \frac{1}{\sqrt{p \lambda_K}} + \frac{n}{p \lambda_K} \right)}\\
	&\times \underbrace{\left( \frac{1}{p}\sum\limits_{g=1}^p\left( \bm{\ell}_g + \frac{1}{n^{1/2}}\tilde{\bm{e}}_{g,1} \right)^T \left( \bm{\ell}_g + \frac{1}{n^{1/2}}\tilde{\bm{e}}_{g,1} \right) \right)^{1/2}}_{\mathcal{O}_P(1)}\\
	& = \mathcal{O}_P\left( \frac{1}{\sqrt{p \lambda_K}} + \frac{n}{p \lambda_K} \right)
	\end{align*}
		\item \begin{equation*}
		\frac{1}{np}\sum\limits_{g=1}^p  \bm{y}_g^T \tilde{\bm{C}}\bm{\delta}^T \bm{\delta}\tilde{\bm{C}}^T \bm{y}_g = \frac{1}{p}\sum\limits_{g=1}^p \left( \bm{\ell}_g + \frac{1}{n^{1/2}}\tilde{\bm{e}}_{g,1} \right)^T \bm{\delta}^T \bm{\delta} \left( \bm{\ell}_g + \frac{1}{n^{1/2}}\tilde{\bm{e}}_{g,1} \right) = o_P\left( \frac{1}{\sqrt{p \lambda_K}} + \frac{n}{p \lambda_K} \right)
		\end{equation*}				

		\end{enumerate}
	
	\end{enumerate}

\item \begin{equation*}
\frac{1}{n-K} \bm{y}_g^T \bm{Q} \hat{\bm{Z}} \hat{\bm{Z}}^T \bm{Q}^T\bm{y}_g = \frac{1}{n-K}\tilde{\bm{e}}_{g,2}^T \hat{\bm{Z}} \hat{\bm{Z}}^T \tilde{\bm{e}}_{g,2} \leq \mathcal{O}_P\left( \frac{n}{p \lambda_K} \right) \frac{1}{n-K}\tilde{\bm{e}}_{g,2}^T \tilde{\bm{e}}_{g,2}
\end{equation*}
	\begin{enumerate}
	\item \begin{equation*}
	\frac{1}{n-K} \bm{y}_g^T \bm{Q} \hat{\bm{Z}} \hat{\bm{Z}}^T \bm{Q}^T\bm{y}_g = \mathcal{O}_P\left( \frac{n}{p \lambda_K} \right) \mathcal{O}_P(1) = \mathcal{O}_P\left( \frac{n}{p \lambda_K} \right)
	\end{equation*}
	\item \begin{equation*}
	\frac{1}{(n-K) p}\sum\limits_{g=1}^p \bm{y}_g^T \bm{Q} \hat{\bm{Z}} \hat{\bm{Z}}^T \bm{Q}^T\bm{y}_g \leq \mathcal{O}_P\left( \frac{n}{p \lambda_K} \right) \frac{1}{p}\sum\limits_{g=1}^p \frac{1}{n-K}\tilde{\bm{e}}_{g,2}^T \tilde{\bm{e}}_{g,2} = \mathcal{O}_P\left( \frac{n}{p \lambda_K} \right)
\end{equation*}
	\end{enumerate}
	
\item \begin{align*}
\frac{1}{n} \bm{y}_g^T \tilde{\bm{C}} \hat{\bm{V}}\hat{\bm{Z}}^T \bm{Q}^T\bm{y}_g &= \left( \bm{\ell}_g + \frac{1}{n^{1/2}}\tilde{\bm{e}}_{g,1} \right)^T \hat{\bm{V}}\hat{\bm{Z}}^T \frac{1}{n^{1/2}} \tilde{\bm{e}}_{g,2}\\
& = \left( \bm{\ell}_g + \frac{1}{n^{1/2}}\tilde{\bm{e}}_{g,1} \right)^T \bm{V}\hat{\bm{Z}}^T \frac{1}{n^{1/2}} \tilde{\bm{e}}_{g,2} + \left( \bm{\ell}_g + \frac{1}{n^{1/2}}\tilde{\bm{e}}_{g,1} \right)^T \bm{\epsilon}\hat{\bm{Z}}^T \frac{1}{n^{1/2}} \tilde{\bm{e}}_{g,2}
\end{align*}	
	\begin{enumerate}
	\item \begin{enumerate}
	\item \begin{align*}
	\abs{\left( \bm{\ell}_g + \frac{1}{n^{1/2}}\tilde{\bm{e}}_{g,1} \right)^T \bm{V}\hat{\bm{Z}}^T \frac{1}{n^{1/2}} \tilde{\bm{e}}_{g,2}}& \leq \norm{\left( \bm{\ell}_g + \frac{1}{n^{1/2}}\tilde{\bm{e}}_{g,1} \right)^T \bm{V}}_2 \norm{\hat{\bm{Z}}^T}_2 \norm{\frac{1}{n^{1/2}} \tilde{\bm{e}}_{g,2}}_2\\
	& = \mathcal{O}_P\left( 1 \right) \mathcal{O}_P\left( \sqrt{\frac{n}{p \lambda_K}} \right) \mathcal{O}_P\left( 1 \right) = \mathcal{O}_P\left( \sqrt{\frac{n}{p \lambda_K}} \right)
	\end{align*}
	\item By the same logic as above, since $\norm{\hat{\bm{Z}}}_2 = \mathcal{O}_P\left( \sqrt{\frac{n}{p \lambda_K}} \right)$ and $\norm{\epsilon}_2 = \mathcal{O}_P\left( \frac{n}{p \lambda_K} \right)$, \begin{equation*}
	\left( \bm{\ell}_g + \frac{1}{n^{1/2}}\tilde{\bm{e}}_{g,1} \right)^T \bm{\epsilon}\hat{\bm{Z}}^T \frac{1}{n^{1/2}} \tilde{\bm{e}}_{g,2} = \mathcal{O}_P\left( \sqrt{\frac{n}{p \lambda_K}}\frac{n}{p \lambda_K} \right) = o_p\left( \sqrt{\frac{n}{p \lambda_K}} \right)
	\end{equation*}
	\end{enumerate}		
	
	\item \begin{enumerate}
	\item \begin{align*}
	\abs{\frac{1}{p}\sum\limits_{g=1}^p \left( \bm{\ell}_g + \frac{1}{n^{1/2}}\tilde{\bm{e}}_{g,1} \right)^T \bm{\epsilon}\hat{\bm{Z}}^T \frac{1}{n^{1/2}} \tilde{\bm{e}}_{g,2}}& \leq \underbrace{\left( \frac{1}{p}\sum\limits_{g=1}^p \left( \bm{\ell}_g + \frac{1}{n^{1/2}}\tilde{\bm{e}}_{g,1} \right)^T \bm{\epsilon}\bm{\epsilon}^T\left( \bm{\ell}_g + \frac{1}{n^{1/2}}\tilde{\bm{e}}_{g,1} \right) \right)^{1/2}}_{\mathcal{O}_P\left( \frac{n}{p \lambda_K} \right)} \\
	&\times \underbrace{\left( \frac{1}{p}\sum\limits_{g=1}^p \frac{1}{n}\tilde{\bm{e}}_{g,2}^T\hat{\bm{Z}}\hat{\bm{Z}}^T\tilde{\bm{e}}_{g,2} \right)^{1/2}}_{\mathcal{O}_P\left( \sqrt{\frac{n}{p \lambda_K}} \right)}\\
	& = o_P\left( \frac{n}{p \lambda_K} \right)
	\end{align*}
	
	\item \begin{align*}
	\left( \bm{\ell}_g + \frac{1}{n^{1/2}}\tilde{\bm{e}}_{g,1} \right)^T \bm{V}\hat{\bm{Z}}^T \frac{1}{n^{1/2}} \tilde{\bm{e}}_{g,2} &= \sqrt{\frac{p}{n}}\begin{pmatrix}
	d_1 \bm{u}_1[g] & \cdots & d_K \bm{u}_K[g]
	\end{pmatrix} \begin{pmatrix}
	\hat{\bm{z}}_1^T \frac{1}{n^{1/2}} \tilde{\bm{e}}_{g,2}\\
	\vdots\\
	\hat{\bm{z}}_K^T \frac{1}{n^{1/2}} \tilde{\bm{e}}_{g,2}
	\end{pmatrix}\\
	& = \sqrt{\frac{p}{n}}d_1 \bm{u}_1[g] \hat{\bm{z}}_1^T \frac{1}{n^{1/2}} \tilde{\bm{e}}_{g,2} + \cdots + \sqrt{\frac{p}{n}}d_K \bm{u}_K[g] \hat{\bm{z}}_K^T \frac{1}{n^{1/2}} \tilde{\bm{e}}_{g,2}
	\end{align*}
	and
	\begin{equation*}
	\frac{1}{p}\sum\limits_{g=1}^p \sqrt{\frac{p}{n}}d_k \bm{u}_k[g] \hat{\bm{z}}_k^T \frac{1}{n^{1/2}} \tilde{\bm{e}}_{g,2} = \underbrace{\frac{1}{\sqrt{pn}}d_k \hat{\bm{z}}_k^T}_{=\mathcal{O}_P\left( \frac{1}{p} \right)} \underbrace{\sum\limits_{g=1}^p \frac{\bm{u}_k[g]}{n^{1/2}}\tilde{\bm{e}}_{g,2}}_{\sim \frac{1}{n^{1/2}} N_n\left( 0, \bm{u}_k^T \bm{\Sigma} \bm{u}_K I_n \right) = \mathcal{O}_P\left( 1 \right)} = \mathcal{O}_P\left( \frac{1}{p} \right).
	\end{equation*}		
		
	\end{enumerate}
	\end{enumerate}
\end{enumerate}

This completes the proof.

\end{proof}

\begin{center}
\line(1,0){425}
\end{center}

It remains to prove Lemma \ref{lemma:Omega.bc}, Theorems \ref{theorem:Thm1} and \ref{thms:Thm2} and Corollary \ref{corollary:OmegaInf}. To do so, we return to assuming $\bm{Y}$ is distributed according \eqref{equation:Y.model}. However, we continue to use $\tilde{\bm{E}}_1$, $\tilde{\bm{E}}_2 \in \mathbb{R}^{p \times (n-d-K)}$, $\tilde{\bm{N}}$, $\bm{v}_k$, $\bm{V}$, $\hat{\bm{v}}_k$, $\hat{\bm{V}}$, $\hat{\bm{z}}_k \in \mathbb{R}^{n-d-K}$, $\hat{\bm{Z}} \in \mathbb{R}^{(n-d-K) \times K}$ and $\bm{R} \in \mathbb{R}^{(n-d-K)\times (n-d-K)}$ defined above in Lemmas \ref{lemma:dk_vk} and \ref{lemma:Asyvk.hat} in what follows.

\begin{proof}[of lemma \ref{lemma:Omega.bc} and \eqref{equation:AsymOmega.bc2} in Theorem \ref{thms:Thm2}]

Recall
\begin{align*}
&\hat{\bm{\Omega}}^{(OLS)}_{bc} = \text{diag}\left( \frac{\hat{\lambda}_1}{\hat{\lambda}_1 - \hat{\rho}}, \ldots, \frac{\hat{\lambda}_K}{\hat{\lambda}_K - \hat{\rho}} \right)\left( \hat{\bm{L}}^T \hat{\bm{L}} \right)^{-1}\hat{\bm{L}}^T\bm{Y}_1 =  \\
&\begin{pmatrix}
\frac{\hat{\lambda}_1}{\hat{\lambda}_1 - \hat{\rho}} & &\\
& \ddots &\\
& & \frac{\hat{\lambda}_K}{\hat{\lambda}_K - \hat{\rho}}
\end{pmatrix}
\begin{pmatrix}\frac{\lambda_1}{\hat{\lambda}_1} & &\\
& \ddots &\\
& & \frac{\lambda_K}{\hat{\lambda}_K}
\end{pmatrix} \left[ \underbrace{\left( \bm{L}^T \bm{L} \right)^{-1}\hat{\bm{L}}^T\bm{B}}_{(a)} + \underbrace{\left( \bm{L}^T \bm{L} \right)^{-1}\hat{\bm{L}}^T\bm{L}}_{(b)}\bm{\Omega}^{(OLS)} + \underbrace{\left( \bm{L}^T \bm{L} \right)^{-1}\hat{\bm{L}}^T\bm{E}_1}_{(c)} \right].
\end{align*}
Using what we learned above, we will go through each one of these terms to prove $\hat{\bm{\Omega}}^{(OLS)}_{bc} - \bm{\Omega}^{(OLS)} = o_p\left( n^{-1/2} \right)$. First, item \ref{item:BoundL} of assumption \ref{assumption:Set1} and assumption \ref{assumption:B} imply $n^{1/2} \left( \bm{L}^T \bm{L} \right)^{-1} \bm{L}^T \bm{B} = o(1)$. We then have

\begin{enumerate}[label=(\alph*)]
\item $\bm{M}_a = \left( \bm{L}^T \bm{L} \right)^{-1}\hat{\bm{L}}^T\bm{B} = \left( \tilde{\bm{L}}^T \tilde{\bm{L}} \right)^{-1} \sqrt{\frac{n}{p}} \hat{\bm{\tilde{L}}}^T\bm{B}$. If we define $\sqrt{\frac{n}{p}}\bm{B} = \tilde{\bm{B}}$, then
\begin{align*}
\bm{M}_a[k,] &= \lambda_k^{-1} \hat{\bm{v}}_k^T \left( \tilde{\bm{L}} + \frac{1}{\sqrt{p}}\tilde{\bm{E}}_1\right)^T \tilde{\bm{B}} + \underbrace{\lambda_k^{-1}\hat{\bm{z}}_k^T \frac{1}{\sqrt{p}} \tilde{\bm{E}}_2^T \tilde{\bm{B}}}_{\mathcal{O}_P\left( \frac{n}{\lambda_k p} \norm{\frac{n}{\lambda_k p} \bm{B}^T \bm{B}}_2 \right) = o_P\left( n^{-1/2} \right)}\\
& = \lambda_k^{-1} \hat{\bm{v}}_k^T \tilde{\bm{L}}^T \tilde{\bm{B}} + \underbrace{\hat{\bm{v}}_k^T \frac{1}{\sqrt{\lambda_k p}} \tilde{\bm{E}}_1^T \left( \frac{1}{\sqrt{\lambda_k}}\tilde{\bm{B}}  \right)}_{=\mathcal{O}_P\left( \frac{1}{\sqrt{\lambda_k p}}\right)= o_P\left( n^{-1/2} \right)} + o_P\left( n^{-1/2} \right). 
\end{align*}
Let $\tilde{\bm{L}}_k = \tilde{\bm{L}}[,k]$ and $\hat{\tilde{\bm{L}}}_k = \hat{\tilde{\bm{L}}}[,k]$. For the first term in the above expression, we have
\begin{align*}
\lambda_k^{-1} \hat{\bm{v}}_k^T \tilde{\bm{L}}^T \tilde{\bm{B}} &= \lambda_k^{-1}\hat{\bm{v}}_k[k]\tilde{\bm{L}}_k^T \tilde{\bm{B}} + \left[  \lambda_k^{-1}\hat{\bm{v}}_k[1]\tilde{\bm{L}}_1^T \tilde{\bm{B}} + \cdots +  \lambda_k^{-1}\hat{\bm{v}}_k[k-1]\tilde{\bm{L}}_{k-1}^T \tilde{\bm{B}}\right.
\\&\left. +  \lambda_k^{-1}\hat{\bm{v}}_k[k+1]\tilde{\bm{L}}_{k+1}^T \tilde{\bm{B}} + \cdots +  \lambda_k^{-1}\hat{\bm{v}}_k[K]\tilde{\bm{L}}_K^T \tilde{\bm{B}} \right] \\
&=\hat{\bm{v}}_k[k]\left( \lambda_k^{-1/2}\tilde{\bm{L}}_k \right)^T \left( \lambda_k^{-1/2}\tilde{\bm{B}} \right) + \left[ \mathcal{O}_P\left( \frac{1}{\sqrt{\lambda_k p}} + \frac{n}{p \sqrt{\lambda_k\lambda_1}} \right) \left( \lambda_1^{-1/2}\tilde{\bm{L}}_1 \right)^T \left( \lambda_k^{-1/2}\tilde{\bm{B}} \right) + \cdots \right.\\
& \left. + \mathcal{O}_P\left( \frac{1}{\sqrt{\lambda_k p}} + \frac{n}{p \sqrt{\lambda_k\lambda_K}} \right) \left( \lambda_K^{-1/2}\tilde{\bm{L}}_K \right)^T \left( \lambda_k^{-1/2}\tilde{\bm{B}} \right) \right]
\end{align*}
Since $\left( \lambda_k^{-1/2}\tilde{\bm{L}}_k \right)^T \left( \lambda_k^{-1/2}\tilde{\bm{B}} \right), \mathcal{O}_P\left( \frac{1}{\sqrt{\lambda_k p}} + \frac{n}{p \sqrt{\lambda_k\lambda_s}} \right) = o_P\left( n^{-1/2} \right)$, the above expression is $o_P\left( n^{-1/2} \right)$.

\item $\left( \bm{L}^T \bm{L} \right)^{-1}\hat{\bm{L}}^T\bm{L} = \left( \tilde{\bm{L}}^T \tilde{\bm{L}} \right)^{-1}\hat{\tilde{\bm{L}}}^T\tilde{\bm{L}} = \begin{pmatrix}
\lambda_1^{-1} & &\\
& \ddots &\\
& & \lambda_K^{-1}
\end{pmatrix} \hat{\tilde{\bm{L}}}^T\tilde{\bm{L}}$ where
\begin{align*}
\hat{\tilde{\bm{L}}}^T\tilde{\bm{L}} &= \hat{\bm{V}}^T \left( \tilde{\bm{L}} + \frac{1}{\sqrt{p}}\tilde{\bm{E}}_1 \right)^T \tilde{\bm{L}} + \hat{\bm{Z}}^T \frac{1}{\sqrt{p}}\tilde{\bm{E}}_2^T \tilde{\bm{L}} \\
&= \underbrace{\bm{\epsilon}^T\left( \tilde{\bm{L}} + \frac{1}{\sqrt{p}}\tilde{\bm{E}}_1 \right)^T \tilde{\bm{L}}}_{i.)} + \underbrace{\bm{V}^T\left( \tilde{\bm{L}} + \frac{1}{\sqrt{p}}\tilde{\bm{E}}_1 \right)^T \tilde{\bm{L}}}_{ii.)} + \underbrace{\hat{\bm{Z}}^T \frac{1}{\sqrt{p}}\tilde{\bm{E}}_2^T \left(\tilde{\bm{L}} + \frac{1}{\sqrt{p}}\tilde{\bm{E}}_1 \right)}_{iii.)} + \mathcal{O}_P\left( \frac{n}{p} \right)
\end{align*}

	\begin{enumerate}[label=(\roman*)]
	\item Suppose $\bm{\epsilon} = \begin{pmatrix}
	\bm{\epsilon}_1 & \cdots & \bm{\epsilon}_K
	\end{pmatrix}$ where $\bm{\epsilon}_k$ was defined in lemma \ref{lemma:Asyvk.hat} as $\hat{\bm{v}}_k - \bm{v}_k$. Since $\bm{\epsilon} = \mathcal{O}_P\left( \frac{n}{\lambda_K p} \right)$ and $\frac{1}{\sqrt{p}}\tilde{\bm{E}}_1^T \tilde{\bm{L}} = \mathcal{O}_P\left( \sqrt{\frac{\lambda_1}{p}} \right)$, then $\norm{\left( \tilde{\bm{L}}^T \tilde{\bm{L}} \right)^{-1} \bm{\epsilon}^T \frac{1}{\sqrt{p}}\tilde{\bm{E}}_1^T \tilde{\bm{L}}}_2 = \lambda_K^{-1/2}\mathcal{O}_P\left( \frac{n}{p \lambda_K}\sqrt{\frac{\lambda_1}{\lambda_K p}} \right) = o_P\left( n^{-1/2} \right).$ And
	\begin{equation*}
	\left( \tilde{\bm{L}}^T \tilde{\bm{L}} \right)^{-1}\bm{\epsilon}^T \tilde{\bm{L}}^T \tilde{\bm{L}} = \begin{pmatrix}
	\bm{\epsilon}_1[1] & \frac{\lambda_2}{\lambda_1}\bm{\epsilon}_1[2] & \cdots & \frac{\lambda_K}{\lambda_1}\bm{\epsilon}_1[2]\\
	\vdots & \ddots & \cdots & \vdots\\
	\frac{\lambda_1}{\lambda_K}\bm{\epsilon}_K[1] & \frac{\lambda_2}{\lambda_K}\bm{\epsilon}_K[2] & \cdots & \bm{\epsilon}_K[K]
	\end{pmatrix} \underbrace{=}_{\text{lemma \ref{lemma:Asyvk.hat}}} o_P\left( n^{-1/2} \right)
	\end{equation*}
	Therefore, $\bm{\epsilon}^T\left( \tilde{\bm{L}} + \frac{1}{\sqrt{p}}\tilde{\bm{E}}_1 \right)^T \tilde{\bm{L}} = o_P\left( n^{-1/2} \right)$.
	
	\item $\left( \tilde{\bm{L}}^T\tilde{\bm{L}} \right)^{-1}\bm{V}^T\left( \tilde{\bm{L}} + \frac{1}{\sqrt{p}}\tilde{\bm{E}}_1 \right)^T \tilde{\bm{L}}$
\begin{align*}
\bm{V}^T\left( \tilde{\bm{L}} + \frac{1}{\sqrt{p}}\tilde{\bm{E}}_1 \right)^T \tilde{\bm{L}} &= \bm{V}^T\left( \tilde{\bm{L}} + \frac{1}{\sqrt{p}}\tilde{\bm{E}}_1 \right)^T \left( \tilde{\bm{L}} + \frac{1}{\sqrt{p}}\tilde{\bm{E}}_1 \right) - \bm{V}^T \tilde{\bm{L}}^T \frac{1}{\sqrt{p}}\tilde{\bm{E}}_1 - \rho	 \bm{V}^T + \mathcal{O}_P\left( \frac{1}{\sqrt{p}} \right)\\
& = \diag\left( \tilde{\lambda}_1 - \rho, \ldots, \tilde{\lambda}_K - \rho \right)\bm{V}^T - \bm{V}^T \tilde{\bm{L}}^T \frac{1}{\sqrt{p}}\tilde{\bm{E}}_1 + \mathcal{O}_P\left( \frac{1}{\sqrt{p}} \right)\\
& = \diag\left( \tilde{\lambda}_1 - \rho, \ldots, \tilde{\lambda}_K - \rho \right)\bm{V}^T - \bm{V}^T\begin{bmatrix}
\leftarrow & \mathcal{O}_P\left( \frac{\sqrt{\lambda_1}}{\sqrt{p}} \right) & \rightarrow\\
\vdots & \ddots & \vdots\\
\leftarrow & \mathcal{O}_P\left( \frac{\sqrt{\lambda_K}}{\sqrt{p}} \right) & \rightarrow
\end{bmatrix} + \mathcal{O}_P\left( \frac{1}{\sqrt{p}} \right)\\
& = \diag\left( \lambda_1, \ldots, \lambda_K \right) + \diag\left( \mathcal{O}_P\left( \sqrt{\frac{\lambda_1}{p}} \right), \ldots, \mathcal{O}_P\left( \sqrt{\frac{\lambda_K}{p}} \right) \right)\\
& - \begin{bmatrix}
\leftarrow & \mathcal{O}_P\left( \frac{\sqrt{\lambda_1}}{\sqrt{p}} \right) & \rightarrow\\
\vdots & \ddots & \vdots\\
\leftarrow & \mathcal{O}_P\left( \frac{\sqrt{\lambda_K}}{\sqrt{p}} \right) & \rightarrow
\end{bmatrix} + \mathcal{O}_P\left( \frac{1}{\sqrt{p}} \right)
\end{align*}
Therefore, 
\begin{equation*}
\left( \tilde{\bm{L}}^T\tilde{\bm{L}} \right)^{-1}\bm{V}^T\left( \tilde{\bm{L}} + \frac{1}{\sqrt{p}}\tilde{\bm{E}}_1 \right)^T \tilde{\bm{L}} = I_K + \mathcal{O}_P\left( \frac{1}{\sqrt{\lambda_K p}} \right)
\end{equation*}
	
	\item  $\left( \tilde{\bm{L}}^T \tilde{\bm{L}} \right)^{-1}\hat{\bm{Z}}^T \frac{1}{\sqrt{p}}\tilde{\bm{E}}_2^T \left(\tilde{\bm{L}} + \frac{1}{\sqrt{p}}\tilde{\bm{E}}_1 \right)$
\begin{equation*}
\left( \tilde{\bm{L}}^T \tilde{\bm{L}} \right)^{-1}\hat{\bm{Z}}^T \frac{1}{\sqrt{p}}\tilde{\bm{E}}_2^T \left(\tilde{\bm{L}} + \frac{1}{\sqrt{p}}\tilde{\bm{E}}_1 \right) = \begin{pmatrix}
\frac{1}{\lambda_1 \sqrt{p}}\hat{\bm{z}}_1^T \tilde{\bm{E}}_2 \sum\limits_{k=1}^K d_k \bm{u}_k \bm{v}_k^T\\
\vdots\\
\frac{1}{\lambda_K \sqrt{p}}\hat{\bm{z}}_K^T \tilde{\bm{E}}_2 \sum\limits_{k=1}^K d_k \bm{u}_k \bm{v}_k^T
\end{pmatrix}
\end{equation*}
The largest row (in magnitude) in the above matrix will obviously be the $K^{\text{th}}$ row, so we need only focus on that row. First,
\begin{align*}
\frac{d_1}{\lambda_K \sqrt{p}}\hat{\bm{z}}_K^T \tilde{\bm{E}}_2 \bm{u}_1 &= \underbrace{\frac{d_1 d_K}{\lambda_K^2 p}\bm{u}_K^T \tilde{\bm{E}}_2 \tilde{\bm{E}}_2^T\bm{u}_1}_{\mathcal{O}_P\left( \frac{n}{p \lambda_K} + \frac{n^{1/2}d_1}{\lambda_K^{3/2}p} \right) = o_P\left( n^{-1/2} \right)} + \underbrace{\frac{d_1 d_K}{\lambda_K^3 p}\bm{u}_K^T\tilde{\bm{E}}_2 R \tilde{\bm{E}}_2^T\bm{u}_1}_{\mathcal{O}_P\left( \frac{d_1 n}{\lambda_K p}\frac{n}{p \lambda_K^{3/2}} + \frac{d_1}{\sqrt{\lambda_K p}}\frac{n}{p \lambda_K^2} \right) = o_P\left( n^{-1/2} \right)}\\
& + \underbrace{\mathcal{O}_P\left( \underbrace{\frac{d_1}{\lambda_K^{3/2}}\sqrt{\frac{n}{p}}\left( \left( \frac{n}{\lambda_K p} \right)^{3/2} + \frac{n^{1/2}}{\lambda_K p} \right)}_{\frac{n}{\lambda_K^2 p} \frac{d_1 n}{p} + \frac{d_1}{\sqrt{\lambda_K p}} \frac{n}{\lambda_K^2 p} } \right)}_{=o_P\left( n^{-1/2} \right)}\\
& = o_P\left( n^{-1/2} \right)
\end{align*}
and second,
\begin{equation*}
\frac{d_K}{\lambda_K \sqrt{p}}\hat{\bm{z}}_K^T \tilde{\bm{E}}_2 \bm{u}_K = \mathcal{O}_P\left( \frac{n}{\lambda_K p} \right) = o_P\left( n^{-1/2} \right)
\end{equation*}
Therefore, $\left( \tilde{\bm{L}}^T \tilde{\bm{L}} \right)^{-1}\hat{\bm{Z}}^T \frac{1}{\sqrt{p}}\tilde{\bm{E}}_2^T \left(\tilde{\bm{L}} + \frac{1}{\sqrt{p}}\tilde{\bm{E}}_1 \right) = o_P\left( n^{-1/2} \right)$.	
	
	\end{enumerate}
We have shown that $\left( \bm{L}^T \bm{L} \right)^{-1}\hat{\bm{L}}^T\bm{L} = I_K + o_P\left( n^{-1/2} \right)$.

\item Recall that $\bm{Y}_1 = \bm{Y} \bm{X}^T\left( \bm{X}\bm{X} \right)^{-1}$ and $\bm{Y}_2 = \bm{Y}\bm{A}$ where $\bm{A}^T \bm{X} = \bm{0}_{(n-d) \times d}$. Since the residuals $\bm{E} \sim MN_{p \times n}\left( 0, \bm{\Sigma}, I_n \right)$, $\bm{E}_1 = \bm{E} \bm{X}^T\left( \bm{X}\bm{X} \right)^{-1}$ and $\bm{E}_2 = \bm{E}\bm{A}$ are independent. And since we use $\bm{Y}_2$ to estimate $\hat{\bm{L}}$, $\hat{\bm{L}}$ and $\bm{E}_1$ are independent. (I abuse notation here. $\tilde{\bm{E}}_1$ and $\bm{E}_1$ are different. $\tilde{\bm{E}}_1$ is defined using the second set of data in part 1). Therefore,
\begin{align*}
\left( \bm{L}^T \bm{L} \right)^{-1}\hat{\bm{L}}^T\bm{E}_1 &\sim \frac{1}{\sqrt{p}} \begin{pmatrix}
\lambda_1^{-1/2} & &\\
& \ddots &\\
& & \lambda_K^{-1/2}
\end{pmatrix} MN_{K \times d} \left(\bm{0}, \begin{pmatrix}
\lambda_1^{-1/2} & &\\
& \ddots &\\
& & \lambda_K^{-1/2}
\end{pmatrix} \hat{\tilde{\bm{L}}}^T \bm{\Sigma}  \hat{\tilde{\bm{L}}} \begin{pmatrix}
\lambda_1^{-1/2} & &\\
& \ddots &\\
& & \lambda_K^{-1/2}
\end{pmatrix},\right.\\
& \left. \left( \frac{1}{n}\bm{X}\bm{X}^T \right)^{-1} \right)\\
&=\mathcal{O}_P\left( \frac{1}{\sqrt{\lambda_K p}} \right) = o_P\left( n^{-1/2} \right).
\end{align*}

\end{enumerate}

The above work shows that
\begin{align*}
\left( \bm{L}^T \bm{L} \right)^{-1}\hat{\bm{L}}^T\bm{B} + \left( \bm{L}^T \bm{L} \right)^{-1}\hat{\bm{L}}^T\bm{L}\bm{\Omega}^{(OLS)} + \left( \bm{L}^T \bm{L} \right)^{-1}\hat{\bm{L}}^T\bm{E}_1 = \bm{\Omega}^{(OLS)} + o_P\left( n^{-1/2} \right).
\end{align*}
Our last task is to understand $\frac{\hat{\lambda}_k}{\hat{\lambda}_k - \hat{\rho}} \frac{\lambda_k}{\hat{\lambda}_k}$.
\begin{align*}
\frac{\hat{\lambda}_k}{\hat{\lambda}_k - \hat{\rho}} \frac{\lambda_k}{\hat{\lambda}_k} &= \left( \frac{\hat{\lambda}_k - \hat{\rho}}{\lambda_k} \right)^{-1} \underbrace{=}_{\text{Lemmas \ref{lemma:dk_vk} and \ref{lemma:Asyvk.hat}}} \left( 1 + \frac{\rho - \hat{\rho}}{\lambda_k} + \underbrace{\mathcal{O}_P\left( \frac{1}{\sqrt{\lambda_k p}} + \frac{n}{\lambda_k p} \right)}_{o_P\left( n^{-1/2} \right)} \right)^{-1} \underbrace{=}_{\text{Lemma \ref{lemma:Sigma}}}\\
& \left( 1 + o_P\left( n^{-1/2} \right) \right)^{-1} = 1 + o_P\left( n^{-1/2} \right).
\end{align*}
Therefore,
\begin{align*}
\hat{\bm{\Omega}}^{(OLS)}_{bc} &= \begin{pmatrix}
\frac{\hat{\lambda}_1}{\hat{\lambda}_1 - \hat{\rho}} & &\\
& \ddots &\\
& & \frac{\hat{\lambda}_K}{\hat{\lambda}_K - \hat{\rho}}
\end{pmatrix}
\begin{pmatrix}\frac{\lambda_1}{\hat{\lambda}_1} & &\\
& \ddots &\\
& & \frac{\lambda_K}{\hat{\lambda}_K}
\end{pmatrix} \left[ \left( \bm{L}^T \bm{L} \right)^{-1}\hat{\bm{L}}^T\bm{B} + \left( \bm{L}^T \bm{L} \right)^{-1}\hat{\bm{L}}^T\bm{L}\bm{\Omega}^{(OLS)} + \left( \bm{L}^T \bm{L} \right)^{-1}\hat{\bm{L}}^T\bm{E}_1 \right]\\
& = \bm{\Omega}^{(OLS)} + o_P\left( n^{-1/2} \right).
\end{align*}

\end{proof}

\begin{center}
\line(1,0){425}
\end{center}

\begin{proof}[of proposition \ref{proposition:BiasedOmega}]
This is a simple consequence of lemma \ref{lemma:ell} and of the above proof of lemma \ref{lemma:Omega.bc} and \eqref{equation:AsymOmega.bc2} in Theorem \ref{thms:Thm2}.
\end{proof}

\begin{center}
\line(1,0){425}
\end{center}

We now have the tools to prove the main results, theorems \ref{theorem:Thm1} and \ref{thms:Thm2}. We return to assuming the model for the data is given by \eqref{equation:Y.model}.

\begin{proof}[of theorem \ref{theorem:Thm1} and the rest of theorem \ref{thms:Thm2}]
Define $\bm{e}_{g,1}$ to be the $g^{\text{th}}$ row of $\bm{E}_1$. Then for site $g$,
\begin{equation*}
\hat{\bm{\beta}}_g^{bc} - \bm{\beta}_g = \bm{\Omega}^{(OLS)} \left( \bm{\ell}_g - \hat{\bm{\ell}}_g \right) + \bm{e}_{g,1} + \left( \bm{\Omega}^{(OLS)} - \hat{\bm{\Omega}}^{(OLS)}_{bc} \right) \hat{\bm{\ell}}_g.
\end{equation*}
We know that $n^{1/2}\bm{e}_{g,1} \sim N_d \left( 0, \sigma_g^2 \left( \frac{1}{n} \bm{X}\bm{X}^T \right)^{-1}  \right)$ is independent of 
\begin{align*}
n^{1/2}\bm{\Omega}^{(OLS)}\left( \bm{\ell}_g - \hat{\bm{\ell}}_g \right) \sim N\left( 0, \sigma_g^2 \bm{\Omega}^{(OLS)}\left( \bm{\Omega}^{(OLS)} \right)^T \right) + o_P(1).
\end{align*}
Therefore,
\begin{align*}
n^{1/2}\bm{e}_{g,1} + n^{1/2}\bm{\Omega}^{(OLS)} \left( \bm{\ell}_g - \hat{\bm{\ell}}_g \right) \sim N\left( 0, \sigma_g^2\bm{\Sigma}_{\bm{X}}^{-1} + \sigma_g^2 \bm{\Omega}^{(OLS)} \left( \bm{\Omega}^{(OLS)} \right)^T \right) + o_P(1).
\end{align*}
. Lastly, since $n^{1/2}\left( \bm{\Omega}^{(OLS)} - \hat{\bm{\Omega}}^{(OLS)} \right) \hat{\bm{\ell}}_g = o_P(1)$ and $\hat{\sigma}_g^2 = \sigma_g^2 + o_P(1)$,
\begin{equation*}
\frac{n^{1/2}}{\hat{\sigma}_g} \left( \hat{\bm{\beta}}_g^{bc} - \bm{\beta}_g \right) \sim N_d\left( 0, \bm{\Sigma}_{\bm{X}}^{-1} + \bm{\Omega}^{(OLS)}\left( \bm{\Omega}^{(OLS)} \right)^T \right) + o_P(1).
\end{equation*}
\end{proof}

\begin{center}
\line(1,0){425}
\end{center}

Next, we use standard multivariate techniques to prove Corollary \ref{corollary:OmegaInf}.
\begin{proof}[of Corollary \ref{corollary:OmegaInf}]
Under the null hypothesis that $\bar{\bm{\Omega}} = 0$, we define
\begin{align*}
\hat{\bar{\bm{\Omega}}} &= \left( \bm{X}^T \bm{X} \right)^{-1} \bm{X}^T \bar{\bm{C}}\\
&= \left( \frac{1}{n}\bm{X}^T \bm{X} \right)^{-1} \frac{1}{n}\sum\limits_{i=1}^n \bm{x}_i \bm{\xi}_i^T
\end{align*}
and let
\begin{align*}
\hat{\bm{s}}_n = \frac{1}{\sqrt{n}} \sum\limits_{i=1}^n \bm{x}_i \bm{\xi}_i^T.
\end{align*}
Define $\bm{a} = \text{vec}\left( \bm{1}_d \times \bm{\xi}_1 \right)$, where $\bm{1}_d \in \mathbb{R}^d$ is the vector of all ones, and $\varphi_{\bm{a}}\left( \bm{t} \right)$, $\bm{t} \in \mathbb{R}^{dK \times dK}$, to be the characteristic function of $\bm{a}$. Under the null hypothesis, the gradient of $\varphi_{\bm{a}}\left( \bm{t} \right)$ is $\bm{0}$ and the Hessian is $-\bm{1}_{d \times d} \otimes \bar{\bm{\Psi}}$, where $\bm{1}_{d \times d} \in \mathbb{R}^{d \times d}$ is the matrix of all ones. Lastly, let $\bm{t} = \left( \bm{t}_1^T, \ldots, \bm{t}_d^T \right)^T$, $\bm{t}_j \in \mathbb{R}^{K}$. If the magnitude of the entries of $\bm{X}$ are bounded above by $x$, we then have that
\begin{align*}
\log \varphi_{\text{vec}\left( \hat{\bm{s}}_n \right)}\left( \bm{t} \right) &= \sum\limits_{i=1}^n \log \varphi_{\bm{a}}\left(\frac{1}{\sqrt{n}} \begin{bmatrix}
x_i[1]\bm{t}_1\\
\vdots\\
x_i[d]\bm{t}_d
\end{bmatrix} \right)\\
&= \sum\limits_{i=1}^n \left\lbrace-\frac{1}{2n}\bm{t}^T \left[\left( \bm{x}_i \bm{x}_i^T\right) \otimes \bar{\bm{\Psi}}\right] \bm{t} + o(\frac{1}{n} x^2 \norm{\bm{t}}_2^2)\right\rbrace\\
&= -\frac{1}{2}\bm{t}^T \left(\bm{\Sigma}_{X} \otimes \bar{\bm{\Psi}}\right) \bm{t} + o(1).
\end{align*}
Therefore,
\begin{align*}
\left( \bm{X}^T \bm{X} \right)^{1/2}\hat{\bar{\bm{\Omega}}} \stackrel{\mathcal{D}}{\to} MN_{d \times K}\left( \bm{0}, I_d, \bar{\bm{\Psi}} \right)
\end{align*}
meaning
\begin{align*}
\left( \bm{X}^T \bm{X} \right)^{1/2}\bm{\Omega}^{(OLS)} &= \left( \bm{X}^T \bm{X} \right)^{1/2}\hat{\bar{\bm{\Omega}}} \left( \frac{1}{n-d} \bar{\bm{\Xi}}^T P_{X}^{\perp}\bar{\bm{\Xi}} \right)^{-1/2}\\
& \stackrel{\mathcal{D}}{\to} MN_{d \times K}\left( \bm{0}, I_d, I_K \right)
\end{align*}
because $\frac{1}{n-d} \bar{\bm{\Xi}}^T P_{X}^{\perp}\bar{\bm{\Xi}} \stackrel{P}{\to} \bar{\bm{\Psi}}$. The result follows from Theorem \ref{thms:Thm2} and an application of Slutsky's Theorem.
\end{proof}

\begin{center}
\line(1,0){425}
\end{center}

The remaining two technical lemmas are used in the proof of lemma \ref{lemma:Asyvk.hat}. For these two lemmas, we assume $\bm{Y}$ is distributed according to \eqref{equation:newY.model} (as it is in Lemmas \ref{lemma:dk_vk} and \ref{lemma:Asyvk.hat}).

\begin{lemma}
\label{lemma:UtU}
Let $\bm{U} = \begin{pmatrix}
\bm{u}_1 & \cdots & \bm{u}_K
\end{pmatrix}$, $\bm{V} = \begin{pmatrix}
\bm{v}_1 & \cdots & \bm{v}_K
\end{pmatrix}$, $\bm{D} = \text{diag}\left( d_1, \ldots, d_K \right)$ and $\tilde{\bm{N}}$ be as defined in lemmas \ref{lemma:dk_vk} and \ref{lemma:Asyvk.hat} and suppose $\frac{n}{p}\bm{L}_s^T \bm{\Sigma}\bm{L}_k = \mathcal{O}\left( \lambda_{k} \right)$ for $s \leq k$. Then
\begin{equation*}
\bm{u}_s^T \bm{\Sigma} \bm{u}_k = \mathcal{O}_P\left( \sqrt{\frac{\lambda_k}{\lambda_s}} \right).
\end{equation*}
\end{lemma}

\begin{proof}
We need to understand how
\begin{equation*}
\bm{U}^T \bm{\Sigma} \bm{U} = \bm{D}^{-1}\bm{V}^T \tilde{\bm{N}}^T\bm{\Sigma} \tilde{\bm{N}}\bm{V} \bm{D}^{-1}
\end{equation*}
behaves. First, let $\bm{R}_i\bm{R}_i^T = \tilde{\bm{L}}^T\bm{\Sigma}^i\tilde{\bm{L}}$ for $i = 1, 2$. Then $\bm{R}_i = \begin{bmatrix}
O\left( \sqrt{\lambda_1} \right) & 0 & \cdots & 0\\
O\left( \sqrt{\lambda_2} \right) & O\left( \sqrt{\lambda_2} \right) & \cdots & 0\\
\vdots & \vdots & \ddots & \vdots\\
O\left( \sqrt{\lambda_K} \right) & O\left( \sqrt{\lambda_K} \right) & \cdots & O\left( \sqrt{\lambda_K} \right)
\end{bmatrix}$ and
\begin{equation*}
\tilde{\bm{N}}^T \bm{\Sigma} \tilde{\bm{N}} = \tilde{\bm{L}}^T \bm{\Sigma}\tilde{\bm{L}} + \frac{1}{\sqrt{p}}\tilde{\bm{L}}^T\bm{\Sigma} \tilde{\bm{E}}_1 + \frac{1}{\sqrt{p}} \tilde{\bm{E}}_1^T \bm{\Sigma}\tilde{\bm{L}} + \underbrace{\gamma}_{=\frac{\Tr\left( \bm{\Sigma}^2 \right)}{p}}I_K + \mathcal{O}_P\left( \frac{1}{\sqrt{p}} \right)
\end{equation*}
The next quantity we need to determine is $\bm{V}\bm{D}^{-1}$:
\begin{equation*}
\bm{V}\bm{D}^{-1} = \bm{D}^{-1} + \begin{bmatrix}
\mathcal{O}_P\left( \frac{1}{\lambda_1^{3/2}p} \right) & \mathcal{O}_P\left( \frac{1}{\sqrt{\lambda_1 \lambda_2 p}} \right) & \cdots & \mathcal{O}_P\left( \frac{1}{\sqrt{\lambda_1 \lambda_K p}} \right)\\
\mathcal{O}_P\left( \frac{1}{\lambda_1 p} \right) & \mathcal{O}_P\left( \frac{1}{\lambda_2^{3/2}p} \right) & \cdots & \mathcal{O}_P\left( \frac{1}{\sqrt{\lambda_2 \lambda_K p}} \right)\\
\vdots & \vdots & \ddots & \vdots\\
\mathcal{O}_P\left( \frac{1}{\lambda_1 p} \right) & \mathcal{O}_P\left( \frac{1}{\lambda_2 p} \right) & \cdots & \mathcal{O}_P\left( \frac{1}{\lambda_K^{3/2}p} \right)
\end{bmatrix} = \bm{D}^{-1} + \bm{e}
\end{equation*}
and
\begin{equation*}
\bm{R}_i^T\bm{V}\bm{D}^{-1} = \mathcal{O}_P(1) + \mathcal{O}_P\left( \frac{1}{\sqrt{\lambda_K p}} \right).
\end{equation*}
Then,
\begin{enumerate}
\item \begin{equation*}
\frac{1}{\sqrt{p}} \tilde{\bm{E}}_1^T \bm{\Sigma}\tilde{\bm{L}}\bm{V}\bm{D}^{-1} \sim \frac{1}{\sqrt{p}} \underbrace{\bm{M}}_{\sim MN_{K \times K}\left( \bm{0}, I_K, I_K \right)} \bm{R}_2^T \bm{V}\bm{D}^{-1} = \mathcal{O}_P\left( \frac{1}{\sqrt{p}} \right)
\end{equation*}
\item \begin{align*}
\bm{D}^{-1}\bm{V}^T \left(	\tilde{\bm{L}}^T \bm{\Sigma}\tilde{\bm{L}} + \gamma I_K\right) \bm{V}\bm{D}^{-1} &= \bm{D}^{-1}\left(	\tilde{\bm{L}}^T \bm{\Sigma}\tilde{\bm{L}} + \gamma I_K\right)\bm{D}^{-1} + \bm{e}^T \left(\tilde{\bm{L}}^T \bm{\Sigma}\tilde{\bm{L}} + \gamma I_K\right)\bm{D}^{-1}\\
& + \bm{D}^{-1}\left(	\tilde{\bm{L}}^T \bm{\Sigma}\tilde{\bm{L}} + \gamma I_K\right) \bm{e}+ \underbrace{\bm{e}^T\left(	\tilde{\bm{L}}^T \bm{\Sigma}\tilde{\bm{L}} + \gamma I_K\right)\bm{e}}_{\bm{e}^T\bm{R}_1\bm{R}_1^T \bm{e} = \mathcal{O}_P\left( \frac{1}{\lambda_K p} \right)}
\end{align*}
	\begin{enumerate}
	\item \begin{equation*}
	\bm{D}^{-1}\left(	\tilde{\bm{L}}^T \bm{\Sigma}\tilde{\bm{L}} + \gamma I_K\right) \bm{e} = \underbrace{\bm{D}^{-1}\bm{R}_1}_{\mathcal{O}_P(1)} \underbrace{\bm{R}_1^T \bm{e}}_{\mathcal{O}_P\left( \frac{1}{\sqrt{\lambda_K p}} \right)} + \mathcal{O}_P\left( \frac{1}{\sqrt{\lambda_K p}} \right)
	\end{equation*}
	\item $\bm{B} = \bm{D}^{-1}\left( \underbrace{\tilde{\bm{L}}^T \bm{\Sigma}\tilde{\bm{L}}}_{=\bm{A}} + \gamma I_K\right)\bm{D}^{-1}$ is such that for $s \leq k$,
	\begin{equation*}
	\bm{B}_{sk} = \frac{\bm{A}_{sk} + \gamma \delta_{sk}}{d_s d_k} \underbrace{=}_{\tilde{\bm{L}}_s^T \bm{\Sigma}\tilde{\bm{L}}_k = \mathcal{O}\left( \lambda_{k} \right)} \mathcal{O}\left( \frac{\lambda_k}{d_s d_k} \right) + \frac{\gamma}{d_s d_k}\delta_{sk} = \mathcal{O}_P\left( \sqrt{\frac{\lambda_k}{\lambda_s}} \right)
	\end{equation*}
	\end{enumerate}
\end{enumerate}
Therefore, for $s \leq k$
\begin{equation*}
\Rightarrow \left[ \bm{U}^T \bm{\Sigma} \bm{U} \right]_{sk} = \mathcal{O}_P\left( \sqrt{\frac{\lambda_k}{\lambda_s}} + \frac{1}{\sqrt{\lambda_K p}} \right) = \mathcal{O}_P\left( \sqrt{\frac{\lambda_k}{\lambda_s}} \right)
\end{equation*}

\end{proof}

\begin{center}
\line(1,0){425}
\end{center}

\begin{lemma}
\label{lemma:RandomR}
Let $\bm{a}_1, \bm{a}_2 \in \mathbb{R}^p$ be linearly independent unit vectors independent of $\tilde{\bm{E}}_2 \sim MN_{p \times (n-K)}\left( 0, \bm{\Sigma}, I_{n-K} \right)$ for $K$ is a fixed constant. Recall from \eqref{equation:RandomR} that $\bm{R} = \frac{1}{p}\tilde{\bm{E}}_2^T \tilde{\bm{E}}_2 - \rho I_{n-K}$ where $\rho = \frac{1}{p}\Tr(\bm{\Sigma})$. Then
\begin{equation*}
\frac{1}{p}\bm{a}_1^T \tilde{\bm{E}}_2 \bm{R} \tilde{\bm{E}}_2^T \bm{a}_2 = \mathcal{O}_P\left( \left( \frac{n}{p} \right)^{2} + \frac{n}{p} \frac{1}{\sqrt{p}} \right).
\end{equation*}
\end{lemma}

\begin{proof}
Since $K$ is a fixed constant not dependent on $n$ or $p$, I will assume $\tilde{\bm{E}}_2 \sim MN_{p \times n}\left( 0, \bm{\Sigma}, I_n \right)$ for notational convenience.
\begin{equation*}
\frac{1}{p}\bm{a}_1^T\tilde{\bm{E}}_2 \bm{R} \tilde{\bm{E}}_2^T \bm{a}_2 = \frac{1}{p^2}\bm{a}_1^T\tilde{\bm{E}}_2 \tilde{\bm{E}}_2^T \tilde{\bm{E}}_2 \tilde{\bm{E}}_2^T \bm{a}_2 - \frac{\rho}{p}\bm{a}_1^T\tilde{\bm{E}}_2 \tilde{\bm{E}}_2^T \bm{a}_2
\end{equation*}
We will focus our efforts on understanding $\frac{1}{p^2}\bm{a}_1^T\tilde{\bm{E}}_2 \tilde{\bm{E}}_2^T \tilde{\bm{E}}_2 \tilde{\bm{E}}_2^T \bm{a}_2$. Define $\bm{A} = \begin{pmatrix}
\bm{a}_1 & \bm{a}_2
\end{pmatrix}$, $\tilde{\bm{A}} = \bm{\Sigma} \bm{A}$ and $\bm{Q} \in \mathbb{R}^{p \times (p-2)}$ s.t. $\bm{A}^T\bm{\Sigma} \bm{Q} = 0_{2 \times (p-2)}$. Let $P_{\tilde{\bm{A}}} = GG^T$ where $G \in \mathbb{R}^{p \times 2}$ and $P_{\tilde{\bm{A}}}^{\perp} = \bm{Q}\bm{Q}^T$. Since $P_{\tilde{\bm{A}}} + P_{\tilde{\bm{A}}}^{\perp} = I_p$, we have
\begin{align*}
\frac{1}{p^2}\bm{a}_1^T\tilde{\bm{E}}_2 \tilde{\bm{E}}_2^T \tilde{\bm{E}}_2 \tilde{\bm{E}}_2^T \bm{a}_2 &= \frac{1}{p^2}\bm{a}_1^T\tilde{\bm{E}}_2 \tilde{\bm{E}}_2^T \left( P_{\tilde{\bm{A}}} + P_{\tilde{\bm{A}}}^{\perp} \right) \tilde{\bm{E}}_2 \tilde{\bm{E}}_2^T \bm{a}_2\\
& = \frac{1}{p^2}\bm{a}_1^T\tilde{\bm{E}}_2 \tilde{\bm{E}}_2^T P_{\tilde{\bm{A}}} \tilde{\bm{E}}_2 \tilde{\bm{E}}_2^T \bm{a}_2 + \frac{1}{p^2}\bm{a}_1^T\tilde{\bm{E}}_2 \tilde{\bm{E}}_2^T P_{\tilde{\bm{A}}}^{\perp} \tilde{\bm{E}}_2 \tilde{\bm{E}}_2^T \bm{a}_2
\end{align*}
Since $\bm{a}_i^T \bm{\Sigma} \bm{a}_i \leq c$ and $\norm{G^T \bm{\Sigma} G}_2 \leq c$, $\norm{\tilde{\bm{E}}_2^T \bm{a}_i}_2 \sim \norm{MN_{n \times 1}\left( 0, I_n, \bm{a}_i^T \bm{\Sigma} \bm{a}_i \right)}_2 = \mathcal{O}_P\left( n^{1/2} \right)$ and $\norm{\tilde{\bm{E}}_2^T G}_2 \sim \norm{MN_{n \times 2}\left( 0, I_n, G^T \bm{\Sigma} G\right)}_2 = \mathcal{O}_P\left( n^{1/2} \right)$. Then By Cauchy-Schwartz,
\begin{equation*}
\frac{1}{p^2}\bm{a}_1^T\tilde{\bm{E}}_2 \tilde{\bm{E}}_2^T P_{\tilde{\bm{A}}} \tilde{\bm{E}}_2 \tilde{\bm{E}}_2^T \bm{a}_2 = \frac{1}{p^2}\underbrace{\bm{a}_1^T\tilde{\bm{E}}_2}_{1 \times n} \underbrace{\tilde{\bm{E}}_2^T G}_{n \times 2} G^T \tilde{\bm{E}}_2 \tilde{\bm{E}}_2^T \bm{a}_2 = \mathcal{O}_P\left( \frac{n}{p} \right)\mathcal{O}_P\left( \frac{n}{p} \right) = \mathcal{O}_P\left( \left( \frac{n}{p} \right)^2 \right)
\end{equation*}
By Craig's Theorem, $\tilde{\bm{E}}^T \bm{a}_i$ and $\tilde{\bm{E}}^T \bm{Q}$ are independent, since $\bm{a}_i^T \bm{\Sigma} \bm{Q} = 0$. We then have
\begin{equation*}
\frac{1}{p^2}\bm{a}_1^T\tilde{\bm{E}}_2 \tilde{\bm{E}}_2^T P_{\tilde{\bm{A}}}^{\perp} \tilde{\bm{E}}_2 \tilde{\bm{E}}_2^T \bm{a}_2 = \frac{1}{p^2}\bm{a}_1^T\tilde{\bm{E}}_2 \tilde{\bm{E}}_2^T \bm{Q}\bm{Q}^T \tilde{\bm{E}}_2 \tilde{\bm{E}}_2^T \bm{a}_2
\end{equation*}
Let $B = \bm{\Sigma}^{1/2}\bm{Q}\bm{Q}^T \bm{\Sigma}^{1/2}$ and let $H \bm{\Delta} H^T$ be its singular value decomposition. Note that $\max \bm{\Delta} \leq c$ since $\bm{Q}\bm{Q}^T$ is just a projection matrix. Therefore, $\tilde{\bm{E}}_2^T \bm{Q}\bm{Q}^T \tilde{\bm{E}}_2 \sim \bm{J}^T \bm{J}$, where $\bm{J} \sim MN_{p \times n}\left( 0, \bm{\Delta}, I_n \right)$ and is independent of $\frac{1}{\sqrt{p}}\tilde{\bm{E}}_2^T \bm{a}_i = \tilde{\bm{a}}_i \in \mathbb{R}^{n \times 1}$. Note that $\norm{\tilde{\bm{a}}_i}_2 = \mathcal{O}_P\left( \sqrt{\frac{n}{p}} \right)$. Define $\delta = \frac{\Tr\left( \bm{\Delta} \right)}{p} = \rho + \mathcal{O}\left( \frac{1}{p} \right)$, $\gamma = \frac{\Tr\left( \bm{\Delta}^2 \right)}{p}$ and $\bm{b}_i = \frac{1}{\norm{\tilde{\bm{a}}_i}_2}\tilde{\bm{a}}_i$. Then
\begin{equation*}
\frac{1}{p^2}\bm{a}_1^T\tilde{\bm{E}}_2 \tilde{\bm{E}}_2^T \bm{Q}\bm{Q}^T \tilde{\bm{E}}_2 \tilde{\bm{E}}_2^T \bm{a}_2 \sim \norm{\tilde{\bm{a}}_1}_2 \norm{\tilde{\bm{a}}_2}_2 \bm{b}_1^T \frac{1}{p}\bm{J}^T \bm{J} \bm{b}_2 = \norm{\tilde{\bm{a}}_1}_2 \norm{\tilde{\bm{a}}_2}_2 \bm{b}_1^T \begin{pmatrix}
\frac{1}{p}\bm{J}_1^T \bm{J}_1 & \cdots & \frac{1}{p}\bm{J}_1^T \bm{J}_n\\
\vdots & \ddots & \vdots\\
\frac{1}{p}\bm{J}_1^T \bm{J}_n & \cdots & \frac{1}{p}\bm{J}_n^T \bm{J}_n
\end{pmatrix}\bm{b}_2
\end{equation*}
\begin{equation*}
\bm{b}_1^T \begin{pmatrix}
\frac{1}{p}\bm{J}_1^T \bm{J}_1 & \cdots & \frac{1}{p}\bm{J}_1^T \bm{J}_n\\
\vdots & \ddots & \vdots\\
\frac{1}{p}\bm{J}_1^T \bm{J}_n & \cdots & \frac{1}{p}\bm{J}_n^T \bm{J}_n
\end{pmatrix}\bm{b}_2 = \sum\limits_{i=1}^n \bm{b}_1[i]\bm{b}_2[i]\frac{1}{p}\bm{J}_i^T \bm{J}_i + \sum\limits_{i \neq q} \bm{b}_1[i]\bm{b}_2[q]\frac{1}{p}\bm{J}_i^T \bm{J}_q
\end{equation*}
\begin{align*}
\sum\limits_{i=1}^n \bm{b}_1[i]\bm{b}_2[i]\frac{1}{p}\bm{J}_i^T \bm{J}_i &\underbrace{=}_{\bm{X}_i = \frac{1}{p}\bm{J}_i^T \bm{J}_i - \delta} \delta \bm{b}_1^T \bm{b}_2 + \underbrace{\sum\limits_{i=1}^n \bm{b}_1[i]\bm{b}_2[i]\bm{X}_i}_{=\bm{X}}\\
\text{Var}(\bm{X}) &= \sum\limits_{i=1}^n \bm{b}_1[i]^2 \bm{b}_2[i]^2\text{Var}\left( \bm{X}_i \right) = \frac{2\gamma}{p} \sum\limits_{i=1}^n \bm{b}_1[i]^2 \bm{b}_2[i]^2 \leq \frac{2\gamma}{p}
\end{align*}
\begin{equation}
\Rightarrow \sum\limits_{i=1}^n \bm{b}_1[i]\bm{b}_2[i]\frac{1}{p}\bm{J}_i^T \bm{J}_i = \delta \bm{b}_1^T \bm{b}_2 + \mathcal{O}_P\left( \frac{1}{\sqrt{p}} \right) = \rho \bm{b}_1^T \bm{b}_2 + \mathcal{O}_P\left( \frac{1}{\sqrt{p}} \right).
\end{equation}
Note that $\mathbb{E}\left( \sum\limits_{i \neq q} \bm{b}_1[i]\bm{b}_2[q]\frac{1}{p}\bm{J}_i^T \bm{J}_q \right) = 0$, meaning
\begin{align*}
\text{Var}\left( \sum\limits_{i \neq q} \bm{b}_1[i]\bm{b}_2[q]\frac{1}{p}\bm{J}_i^T \bm{J}_q \right) = \mathbb{E}\left[ \left( \sum\limits_{i \neq q} \bm{b}_1[i]\bm{b}_2[q]\frac{1}{p}\bm{J}_i^T \bm{J}_q \right)^2 \right].
\end{align*}
Therefore,
\begin{equation*}
\text{Var}\left( \sum\limits_{i \neq q} \bm{b}_1[i]\bm{b}_2[q]\frac{1}{p}\bm{J}_i^T \bm{J}_q \right) = \frac{1}{p^2}\sum\limits_{i \neq q} \sum\limits_{r \neq s} \bm{b}_1[i]\bm{b}_2[q]\bm{b}_1[r]\bm{b}_2[s] \mathbb{E}\left[ \left( \bm{J}_i^T \bm{J}_q \right)\left( \bm{J}_r^T \bm{J}_s \right) \right].
\end{equation*}
We then need to go through various scenarios to evaluate the above expression.
\begin{enumerate}
\item $i \neq r,s$ and $q \neq r, s$. Then,
\begin{equation*}
\mathbb{E}\left[ \left( \bm{J}_i^T \bm{J}_q \right)\left( \bm{J}_r^T \bm{J}_s \right) \right] = 0
\end{equation*}
\item $i = r$.
	\begin{enumerate}
	\item $q \neq s$
	\begin{equation*}
	\mathbb{E}\left[ \left( \bm{J}_i^T \bm{J}_q \right)\left( \bm{J}_i^T \bm{J}_s \right) \right] = \mathbb{E}\bm{J}_q^T \mathbb{E} \left[ \bm{J}_i \bm{J}_i^T \mid \bm{J}_q, \bm{J}_s \right]  \bm{J}_s = \mathbb{E} \bm{J}_q^T \bm{\Delta} \bm{J}_s = \Tr\left( \bm{\Delta} \mathbb{E}\bm{J}_s \bm{J}_q^T \right) = 0
	\end{equation*}
	\item $q = s$
	\begin{equation*}
	\mathbb{E}\left[ \left( \bm{J}_i^T \bm{J}_q \right)\left( \bm{J}_i^T \bm{J}_q \right) \right] = \mathbb{E}\bm{J}_q^T \mathbb{E} \left[ \bm{J}_i \bm{J}_i^T \mid \bm{J}_q\right]  \bm{J}_q = \mathbb{E} \bm{J}_q^T \bm{\Delta} \bm{J}_q = \Tr \left( \bm{\Delta}^2 \right) = p \gamma
	\end{equation*}
	
	\end{enumerate}
\item $i = s$
	\begin{enumerate}
	\item $q \neq r$
	\begin{equation*}
	\mathbb{E}\left[ \left( \bm{J}_i^T \bm{J}_q \right)\left( \bm{J}_r^T \bm{J}_i \right) \right] = \mathbb{E}\bm{J}_q^T \mathbb{E} \left[ \bm{J}_i \bm{J}_i^T \mid \bm{J}_q, \bm{J}_r \right]  \bm{J}_r = \mathbb{E} \bm{J}_q^T \bm{\Delta} \bm{J}_r = 0
	\end{equation*}
	\item $q = r$
	\begin{equation*}
	\mathbb{E}\left[ \left( \bm{J}_i^T \bm{J}_q \right)\left( \bm{J}_q^T \bm{J}_i \right) \right] = p\gamma
	\end{equation*}
	
	\end{enumerate}
	
\item $q = s$, $i \neq r$ (we already have the case $q = s, i=r$ above).
\begin{equation*}
\mathbb{E}\left[ \left( \bm{J}_i^T \bm{J}_q \right)\left( \bm{J}_r^T \bm{J}_q \right) \right] = 0
\end{equation*}
\item $q = r$, $i \neq s$ (we already have the case $q = r, i=s$ above).
\begin{equation*}
\mathbb{E}\left[ \left( \bm{J}_i^T \bm{J}_q \right)\left( \bm{J}_q^T \bm{J}_s \right) \right] = 0
\end{equation*}
\end{enumerate}
Therefore,
\begin{equation*}
\frac{1}{p^2}\sum\limits_{i \neq q} \sum\limits_{r \neq s} \bm{b}_1[i]\bm{b}_2[q]\bm{b}_1[r]\bm{b}_2[s] \mathbb{E}\left[ \left( \bm{J}_i^T \bm{J}_q \right)\left( \bm{J}_r^T \bm{J}_s \right) \right] = \frac{\gamma}{p}\sum\limits_{i \neq q} \bm{b}_1[i]^2 \bm{b}_2[q]^2 + \frac{\gamma}{p}\sum\limits_{i \neq q} \bm{b}_1[i]\bm{b}_2[i]\bm{b}_1[q]\bm{b}_2[q]
\end{equation*}
\begin{equation*}
\sum\limits_{i \neq q} \bm{b}_1[i]^2 \bm{b}_2[q]^2 = \sum\limits_{i=1}^n \bm{b}_1[i]^2 \sum\limits_{q \neq i}^n \bm{b}_2[q]^2 \leq \sum\limits_{i=1}^n \bm{b}_1[i]^2 \sum\limits_{q=1}^n \bm{b}_2[q]^2 = 1
\end{equation*}
\begin{equation*}
\sum\limits_{i \neq q} \bm{b}_1[i]\bm{b}_2[i]\bm{b}_1[q]\bm{b}_2[q] = \sum\limits_{i=1}^n \bm{b}_1[i]\bm{b}_2[i] \sum\limits_{q \neq i}^n \bm{b}_1[q]\bm{b}_2[q], \quad \abs{\sum\limits_{q \neq i}^n \bm{b}_1[q]\bm{b}_2[q]} \leq \norm{\bm{b}_{1,-i}}_2 \norm{\bm{b}_{2,-i}}_2 \leq 1
\end{equation*}
\begin{equation*}
\Rightarrow \abs{\sum\limits_{i=1}^n \bm{b}_1[i]\bm{b}_2[i] \sum\limits_{q \neq i}^n \bm{b}_1[q]\bm{b}_2[q]} \leq 	\left( \sum\limits_{i=1}^n \left( \sum\limits_{q \neq i}^n \bm{b}_1[q]\bm{b}_2[q] \right)^2 \bm{b}_1[i]^2 \right)^{1/2} \norm{\bm{b}_2}_2 \leq \norm{\bm{b}_1}_2 \norm{\bm{b}_2}_2 = 1
\end{equation*}
Therefore $\text{Var}\left( \sum\limits_{i \neq q} \bm{b}_1[i]\bm{b}_2[q]\frac{1}{p}\bm{J}_i^T \bm{J}_q \right) \leq \frac{\gamma}{p}$, meaning
\begin{align*}
\frac{1}{p^2}\bm{a}_1^T\tilde{\bm{E}}_2 \tilde{\bm{E}}_2^T \tilde{\bm{E}}_2 \tilde{\bm{E}}_2^T \bm{a}_2 &= \norm{\tilde{\bm{a}}_1}_2 \norm{\tilde{\bm{a}}_2}_2\rho \bm{b}_1^T \bm{b}_2 + \norm{\tilde{\bm{a}}_1}_2 \norm{\tilde{\bm{a}}_2}_2 \mathcal{O}_P\left( \frac{1}{\sqrt{p}} \right) + \mathcal{O}_P\left( \left( \frac{n}{p} \right)^2 \right)\\
& = \frac{\rho}{p}\bm{a}_1^T \tilde{\bm{E}}_2 \tilde{\bm{E}}_2^T \bm{a}_2 + \mathcal{O}_P\left( \frac{n}{p}\frac{1}{\sqrt{p}} \right) + \mathcal{O}_P\left( \left( \frac{n}{p} \right)^2 \right)
\end{align*}
\begin{equation*}
\Rightarrow \frac{1}{p}\bm{a}_1^T\tilde{\bm{E}}_2 \bm{R} \tilde{\bm{E}}_2^T \bm{a}_2 = \mathcal{O}_P\left( \frac{n}{p}\frac{1}{\sqrt{p}} \right) + \mathcal{O}_P\left( \left( \frac{n}{p} \right)^2 \right).
\end{equation*}
\end{proof}

\end{document}